\documentclass[10pt,preprint]{aastex}
\usepackage{epsfig}
\usepackage{amsmath}
\usepackage{upgreek}
\usepackage{natbib}
\usepackage{graphicx}
\usepackage{longtable}
\usepackage{url}

\newcommand{\noprint}[1]{}

\begin{document}

\title{Deuterated molecules in Orion KL from Herschel/HIFI\footnote{Herschel is an ESA space observatory with science instruments provided by European-led Principal Investigator consortia and with important participation from NASA.}}

\author{Justin L. Neill, Nathan R. Crockett, Edwin A. Bergin}
\affil{Department of Astronomy, University of Michigan, 500 Church Street, Ann Arbor, MI 48109, USA; jneill@umich.edu}

\author{John C. Pearson}
\affil{Jet Propulsion Laboratory, Caltech, Pasadena, CA 91109, USA}

\author{Li-Hong Xu}
\affil{Department of Physics, Centre for Laser, Atomic, and Molecular Studies (CLAMS), University of New Brunswick, Saint John, New Brunswick, Canada E2L 4L5}


\begin{abstract}

We present a comprehensive study of the deuterated molecules detected in the fullband HIFI survey of the Orion KL region.  Ammonia, formaldehyde, and methanol and their singly deuterated isotopologues are each detected through numerous transitions in this survey with a wide range in optical depths and excitation conditions.  In conjunction with a recent study of the abundance of HDO and H$_2$O in Orion KL, this study yields the best constraints on deuterium fractionation in an interstellar molecular cloud to date.  As previous studies have found, both the Hot Core and Compact Ridge regions within Orion KL contain significant abundances of deuterated molecules, suggesting an origin in cold grain mantles.  In the Hot Core, we find that ammonia is roughly a factor of 2 more fractionated than water.  In the Compact Ridge, meanwhile, we find similar deuterium fractionation in water, formaldehyde, and methanol, with D/H ratios of (2---8) $\times$ $10^{-3}$.  The [CH$_2$DOH]/[CH$_3$OD] ratio in the Compact Ridge is found to be $1.2 \pm 0.3$.  The Hot Core generally has lower deuterium fractionation than the Compact Ridge, suggesting a slightly warmer origin, or a greater contribution from warm gas phase chemistry.

\end{abstract}

\section{Introduction}

Observations of deuterated species in molecular clouds are a powerful diagnostic of chemical evolution through the stages of star formation \citep{millar05}.  The slightly lower zero-point vibrational energy of deuterated molecules relative to their hydrogenated counterparts leads, at low temperatures, to molecular deuterium isotopic ratios that greatly exceed the cosmic abundance of D relative to H ($\sim10^{-5}$).  In the hot molecular cores and corinos associated with the protostellar phase of star formation, the observed gas temperatures ($\sim$100--300 K) are too high for substantial deuterium fractionation, so enhanced D/H ratios are typically interpreted as a fossil of an earlier, colder phase \citep{jacq90, rodgers96, millar03, roberts07}.  Because different molecules have different deuterium enhancement processes, the full characterization of the deuterium fractionation in a star-forming region can provide constraints on its evolutionary history.  Moreover, enhanced molecular D/H ratios are also observed in solar system comets and meteorites, which could be due to an origin in the cold interstellar cloud that collapsed to form the Sun \citep{caselli12}.  Therefore, understanding deuterium fractionation in star-forming regions is important to tracing our interstellar origins.

The Orion Kleinmann-Low nebula (Orion KL), as the brighest region within the Orion molecular cloud, is notable for its rich organic molecular inventory \citep{blake87, blake96, schilke01, comito05, persson07, tercero10}, and several organic species there have been observed to have substantial deuterium fractionation, including water \citep{jacq90, neill13}, ammonia \citep{rodriguezkuiper78, walmsley87}, formaldehyde \citep{loren85, turner90}, and methanol \citep{jacq93, peng12}.  These molecules are some of the most abundant organic species in Orion KL and other molecular clouds, and are likely precursors to much of the molecular complexity of the interstellar medium \citep{garrod08}.  Observations of deuterated molecules can provide clues into their chemistry and so are relevant to understanding the pathways to molecular complexity in star-forming regions.  A number of studies using ground-based observatories have derived D/H ratios for these basic molecules, but suffer from the important limitation that these light molecules have widely spaced transitions, and so ratios are have been based on the observation of only a few transitions.  Because of the complex excitation of molecular lines in Orion KL and the multiple spatial/velocity components in the region, this leads to substantial uncertainties on the total molecular abundances.

Using the Heterodyne Instrument for the Far Infrared (HIFI) \citep{degraauw10} aboard the Herschel Space Observatory \citep{pilbratt10}, a broadband, high spectral resolution (1.1 MHz) far-infrared survey of Orion KL has been acquired as part of the Herschel Observations of EXtra-Ordinary Sources (HEXOS) key program \citep{bergin10}.  Much of this spectral region is inaccessible from the ground due to atmospheric opacity, so this is the first high-resolution survey of Orion KL in much of this wavelength range.  A previous study presented the abundances of HDO and H$_2$O in the various spatial/velocity components with Orion KL: the Hot Core, Compact Ridge, and Plateau \citep{neill13}.  Here we derive abundances and D/H ratios for  ammonia (NH$_2$D), formaldehyde (HDCO), and methanol (CH$_2$DOH and CH$_3$OD).  Each of these molecules have a large number of transitions in the HIFI survey, spanning a wide range in upper-state energies and line strengths.  With the consistent observation strategy, calibration, and high spectral resolution of the survey, this is the most comprehensive investigation of deuterium fractionation in a star-forming region to date.

In \S 2 we provide an overview of the observations used in this study.  Following this, \S 3 describes the modeling and results: \S 3.1 provides an overview of the fitting and modeling approach, and \S 3.2--3.4 describe the results for ammonia, formaldehyde, and methanol, respectively.  \S 4 discusses the derived D/H ratios in the context of current models for deuterium fractionation in star-forming regions, and \S 5 concludes.

\section{Observations}

The HIFI survey of Orion KL used in this work covers the frequency ranges 480--1280, 1426--1535, and 1573--1906 GHz.  All transitions used for this work are in bands 1--5 (480--1280 GHz), with a pointing position of $\alpha_{\textnormal{J2000}} = 05^\textnormal{h} 35^\textnormal{m} 14^\textnormal{s}.3$, $\delta_{\textnormal{J2000}} = -05^\circ 22'33.7''$.  The Herschel beam size ranges from $44''$ at 480 GHz to $16.5''$ at 1280 GHz, so emission from the two primary sources of compact emission in Orion KL, the Hot Core and Compact Ridge, is included, as well as from the more extended Plateau (see further discussion of the spatial/velocity components in \S 3.1).  Further details on the data reduction can be found in Crockett et al. (2013b, in preparation).  Briefly, all spectra were acquired over the period from March 2010 to April 2011 in dual beam switch mode using the wide band spectrometer at 1.1 MHz spectral resolution.  Reference positions lying $3'$ east or west of the science target were used.  All data presented here were processed with HIPE \citep{ott10} version 5.0, using the standard HIFI deconvolution (the \emph{doDeconvolution} task), with the H and V polarizations averaged together to improve the signal to noise ratio.  For bands 1--5, calibration was performed using aperture efficiencies, which can be found in \cite{roelfsema12}.

An Atacama Large Millimeter/Submillimeter Array (ALMA) survey of Orion KL (214--247 GHz) collected as part of its Science Verification was used to create an image of a transition of NH$_2$D at 239.848 GHz and of $^{13}$CH$_3$OH at 235.881 GHz.  The full calibrated measurement set is available at \url{https://almascience.nrao.edu/alma-data/science-verification}.  Further details of the observation can be found in, e.g., \cite{fortman12b, neill13}.  To create the images presented here, the continuum emission as estimated from line-free channels near each transition was subtracted, and the CLEAN algorithm with robust weighting and a Briggs parameter of 0.0 was used for deconvolution.  The spectral resolution of the NH$_2$D image was $1.76'' \times 1.16''$, and $1.66'' \times 0.94''$ for the $^{13}$CH$_3$OH image.  The millimeter continuum map used here is available at the ALMA science verification website, and was created from the same data set using the multi-frequency synthesis CLEAN mode of 30 line-free channels at 230.9 GHz, with a resolution of $1.86'' \times 1.37''$.

\section{Results}

\subsection{Modeling approaches}

Several distinct spatial components have been observed in previous observations toward Orion KL.  These components are typically identified in single-dish spectra by their characteristic kinematic parameters ($v_\textnormal{LSR}$ and $\Delta v$), and have been associated with different spatial regions by interferometric observations \citep{wright96, blake96, beuther05, friedel08}.  Table 1 summarizes the kinematic and physical parameters of the spatial/velocity components in Orion KL based on the numerous previous studies of this region.  Some of the species described here have complex lineshapes due to their detection in more than one of these components.  In these cases, the contributions of different components are separated by summing multiple Gaussians to model the observed lineshapes, enabled by the high spectral resolution of these observations.  The kinematic parameters listed in Table 1 should only be seen as a guideline; different molecules have been observed using interferometric arrays to have different spatial distributions within each of these sources, so these parameters can vary from molecule to molecule as well.

\begin{deluxetable}{c c c c c c c}
\tablenum{1}
\tablecaption{Kinematic parameters and physical conditions of the Orion KL spatial components.\tablenotemark{a}}
\tablewidth{0pt}
\tablehead{Component & $\theta_s$ & $v_\textnormal{LSR}$ & $\Delta v$ & $T_\textnormal{kin}$ & $n$(H$_2$) & $N$(H$_2$) \\
 & $('')$ & (km s$^{-1}$) & (km s$^{-1}$) & (K) & (cm$^{-3})$ & (cm$^{-2})$}
\startdata
Hot Core & 5--10 & 3--5 & 5--10 & 150--300 & $10^7$--$10^8$ & $3.1 \times 10^{23}$ \\
Compact Ridge & 5--15 & 7--9 & 3--5 & 80--125 & $10^6$--$10^7$ & $3.9 \times 10^{23}$ \\
Plateau & 20--30 & 6--12 & 20--25 & 100--150 & $10^6$--$10^7$ & $1.8 \times 10^{23}$ \\
Extended Ridge & 180 & 8--10 & 3--4 & 40--60 & $\sim$$10^5$ & $7.1 \times 10^{22}$ \\
\enddata
\tablenotetext{a}{Values compiled from \cite{blake87, tercero10, melnick10, plume12}, and Crockett et al. (2013b, in preparation).}
\end{deluxetable}

The HIFI spectrum of Orion KL has an extremely high line density \citep{bergin10}, and a full analysis of this spectrum is underway (Crockett et al. 2013b, in preparation).  The fullband model to the spectrum is used in this study to determine which lines are blended with known lines of other molecules.  Where possible, these blends have been accounted for, either by subtracting the line flux predicted by the model, or, if needed, by fitting the blended lines to Gaussian components.  Transitions that are heavily blended, particularly with stronger lines, have been neglected for this analysis.  Transitions of the different species discussed here were modeled using the spectroscopic data compiled in the CDMS \citep{muller01, muller05} and JPL \citep{pickett98} catalogs.  The primary spectroscopic data for deuterated ammonia comes from \cite{muller10, delucia76, cohen82, fusina88}.  For formaldehyde, the line frequencies and intensities used are drawn from \cite{bocquet96} and \cite{brunken03} for H$_2$CO, \cite{muller00} for H$_2$$^{13}$CO, and \cite{bocquet99} and \cite{johns77} for HDCO.  The parameters for $^{13}$CH$_3$OH come from \cite{xu96}, and for CH$_2$DOH from \cite{pearson12}.  The full spectroscopic data set for CH$_3$OD is unpublished, but information about the Hamiltonian model and partial line lists can be found in \cite{walsh00}.

\subsection{Ammonia}

The rotational energy level structure of NH$_2$D has some significant differences from that of NH$_3$, as do the selection rules for radiative and collisional transitions.  Just as for NH$_3$, an ``umbrella'' inversion motion splits each rotational level (denoted by the quantum numbers $J_{K_a K_c}$) into two inversion states, labeled as + and -, corresponding to symmetric and antisymmetric combinations of the two equivalent minimum-energy structures \citep{ho78}; the symmetric (+) configuration is lower in energy by about 12 GHz (0.58 K).  In NH$_3$, radiative rotational transitions between the + and - states for a given $J_K$ state are allowed (inversion transitions), as well as transitions with$\Delta J = 1, \Delta K = 0$ (rotation-inversion transitions).  (Transitions with $\Delta K = 3, 6...$ are also weakly allowed.)  States where $J = K$ are referred to as metastable, in that population in these states is transferred to lower energy levels predominantly through collisions, so these transitions do not require high H$_2$ densities to be excited, leading to the use of NH$_3$ as a common temperature probe.  Transitions involving non-metastable levels ($J > K$), however, require high densities ($\sim$$10^8$ cm$^{-3}$) to be excited, although the populations of these levels are also highly coupled to the far-infrared radiation field \citep{hermsen88b, schilke92}.

NH$_2$D is an asymmetric top molecule with $\mu_a =$ 0.18 D and $\mu_c =$ 1.46 D \citep{cohen82}.  The \emph{ortho} levels are those in the + state with odd $K_a$, and in the - state with even $K_a$, while the remaining states are \emph{para}.  \emph{Ortho} energy levels have an factor of 3 greater spin weights than corresponding levels of \emph{para}.  Transitions between levels with the same $J_{K_a K_c}$ are forbidden because they belong to different nuclear spin states, and so the well-known centimeter transitions that can be used to probe NH$_3$ to high energies do not exist in NH$_2$D.  There are, however, rotational transitions of NH$_2$D throughout the millimeter and submillimeter regions of the spectrum.  In Figure 1, the energy level structure of \emph{ortho}-NH$_2$D is shown, with the transtions detected in the HIFI fullband spectrum toward Orion KL indicated as arrows.  Because both $a$- and $c$-type radiative transitions are allowed, there are no metastable levels for this isotopologue.

\begin{figure*}
\figurenum{1}
\centering
\includegraphics[width=6.5in]{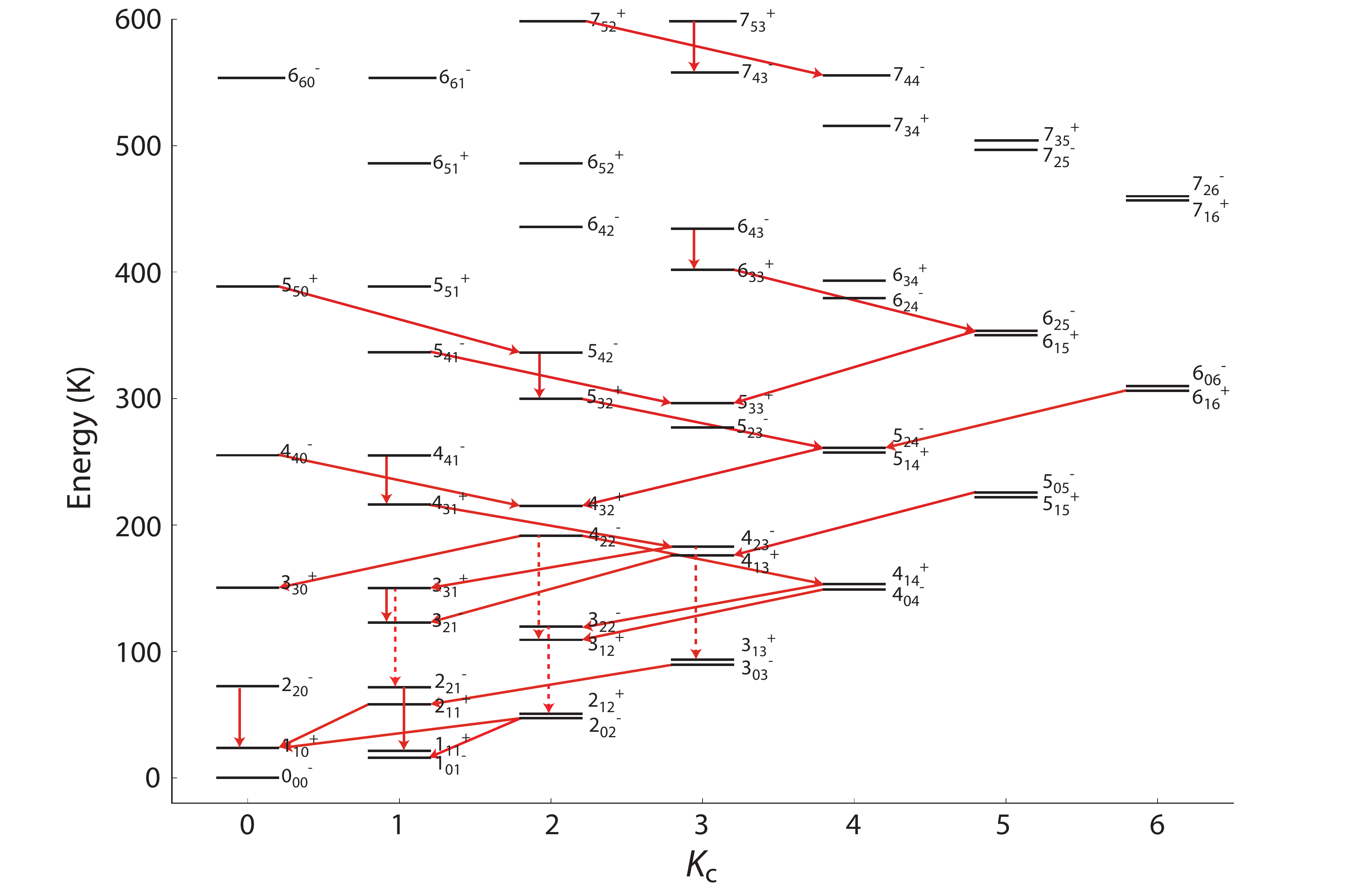}
\caption{Energy levels of \emph{ortho}-NH$_2$D.  The solid red arrows indicate transitions that are cleanly detected toward Orion KL and used for the determination of the NH$_2$D column density.  Dashed arrows indicate lines that are detected in bands 6 and 7 of HIFI, but are not used for the column density determination due to high optical depth.  In cases where two states are closer in energy than can be resolved on the figure's energy scale, the lines have been separated for clarity.}
\end{figure*}

NH$_2$D was first detected toward Orion KL from ground-based observatories through two millimeter transitions ($1_{11}$$^+ -1_{01}$$^-$ and $1_{11}$$^- -1_{01}$$^+$) by \cite{rodriguezkuiper78}, followed by the measurement of two higher-energy lines at centimeter frequencies ($3_{13}$$^+-3_{03}$$^-$ and $4_{14}$$^- -4_{04}$$^+$) by \cite{walmsley87}.  The $1_{11}$$^- -1_{01}$$^+$ transition has also been mapped with an interferometer at $5'' \times 4''$ spatial resolution by \cite{saito94}.  Just as in those studies, NH$_2$D lines in the HIFI spectrum of Orion KL are reasonably well modeled by single Gaussian components with $v_\textnormal{LSR} \sim 6.5$ km s$^{-1}$ and $\Delta v \sim 5$ km s$^{-1}$, suggesting an origin in the Hot Core region.  Some of the lower-energy lines have asymmetric profiles with a ``tail" to bluer velocities that suggests a second component.  Similar lineshapes can be seen in NH$_3$ inversion lines in the Very Large Array measurements of \cite{goddi11}.  None of the lineshapes measured with HIFI are measurably affected by the nuclear quadrupole hyperfine structure induced by the $^{14}$N nucleus, which causes splittings of less than 0.5 km s$^{-1}$ for all measured transitions.  A sample of NH$_2$D lines observed with HIFI is presented in Figure 2.  Table 2 presents the fit Gaussian parameters for all detected lines.  In this Table and the following tables of fit transitions, we neglect any lines that are strongly blended with transitions of other species.

\begin{figure*}
\figurenum{2}
\includegraphics[width=6.0in]{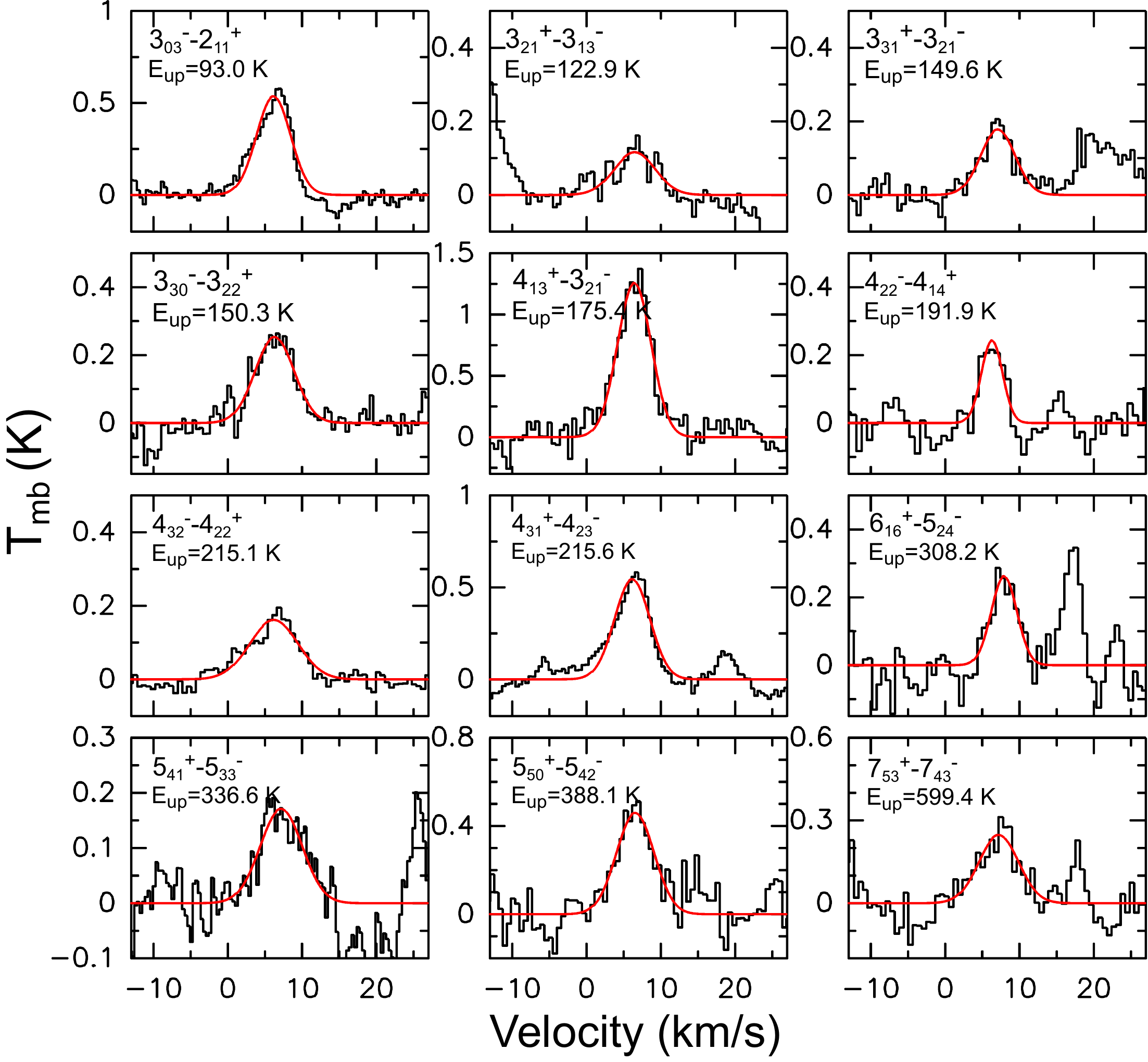}
\caption{A sample of NH$_2$D lines detected in the Orion KL HIFI spectrum.  The single-component Gaussian line fits are shown in red.}
\end{figure*}

NH$_3$ is also detected in the HIFI fullband survey toward Orion KL via eight rotation-inversion transitions.  However, due to the symmetry of NH$_3$, all of these transitions are low in energy ($E_\textnormal{up} < 200$ K) and so their lineshapes are dominated by broad outflow components.  These transitions therefore provide little information on the NH$_3$ column density in the hot core.  The $^{15}$NH$_3$ isotope is also detected, but also has only low-energy transitions in the HIFI bandwidth.  Therefore, we use the centimeter inversion transitions, which have been observed going over a wide range in $J$ and $K$, including both metastable and non-metastable levels, ranging up to states with energies near 2000 K, with the 100 m Effelsberg telescope \citep{hermsen88b, wilson93}.  Because the hyperfine structure is resolved for many of these low-frequency lines, direct determinations of the optical depths of the transitions could be made to derive the NH$_3$ column density.  Three spatial components were detected for NH$_3$ lines in the study of \cite{hermsen88b}:  the Plateau, a ``spike'' component, and the Hot Core.  The spike component has $v_\textnormal{LSR} \sim 8$ km s$^{-1}$ and $\Delta v \sim 2.7$ km s$^{-1}$, consistent with either the Compact Ridge or Extended Ridge.  Three non-metastable inversion transitions were observed to have this velocity component, as well as the higher-energy (409 K) $J_K = 6_6$ transition, suggesting that at least part of this emission arises from warm and dense gas.  The highest-energy NH$_3$ lines are observed to emit from only the Hot Core component.  VLA observations of high-energy NH$_3$ inversion lines have shown that high-energy ammonia emission arises from compact regions in the Hot Core surounding the strongest radio and infrared continuum sources in Orion KL \citep{wilson00, goddi11}.

In Figure 3 we show an integrated flux map of the $3_{22}$$^--3_{12}$$^+$ transition of \emph{ortho}-NH$_2$D ($\nu = 239848.1$ MHz, $E_u = 120$ K) as measured by ALMA.  The $3_{22}$$^+-3_{12}$$^-$ transition is also in the bandwidth of the available ALMA data, but this transition is blended with the $9_{18}-9_{19}$ transition of H$_2$CO,  so we do not show it here.  We observe most of the total flux in the hot core (extending from northeast-to-southwest nearly aligned with the dust continuum emission), with the northern component peaking at a velocity of approximately 3 km s$^{-1}$ and the southern component near 6 km s$^{-1}$.  There is also a peak to the west of the Hot Core nearly coincident with the IRc7 continuum source.  The total distribution of the emission is similar to the high-energy inversion transitions of NH$_3$ \citep{goddi11}.  There is also some similarity to the distribution of HDO emission as measured by the ALMA survey, reported by \cite{neill13}.  In that study, the clump to the south of the Hot Core, with a velocity of 7 km s$^{-1}$, was the brighest emission in the image, and in the HIFI analysis this component was identified as having a very high abundance of HDO, as evidenced by the detection of six transitions of HD$^{18}$O with this velocity, and of H$_2$O ($\chi$(H$_2$O) = $6.5 \times 10^{-4}$).  In this analysis, we treat the Hot Core as a single spatial component.  \cite{hermsen88b} found the hot core component of NH$_3$ to have a spatial extent of 6'', derived from the optically thick low-$J$ inversion lines.  We adopt here the same source size for the determination of the NH$_2$D column density from the HIFI survey; this value is consistent with the area of the emissive clumps seen in Figure 3.

\begin{figure}
\figurenum{3}
\includegraphics[width=6.0in]{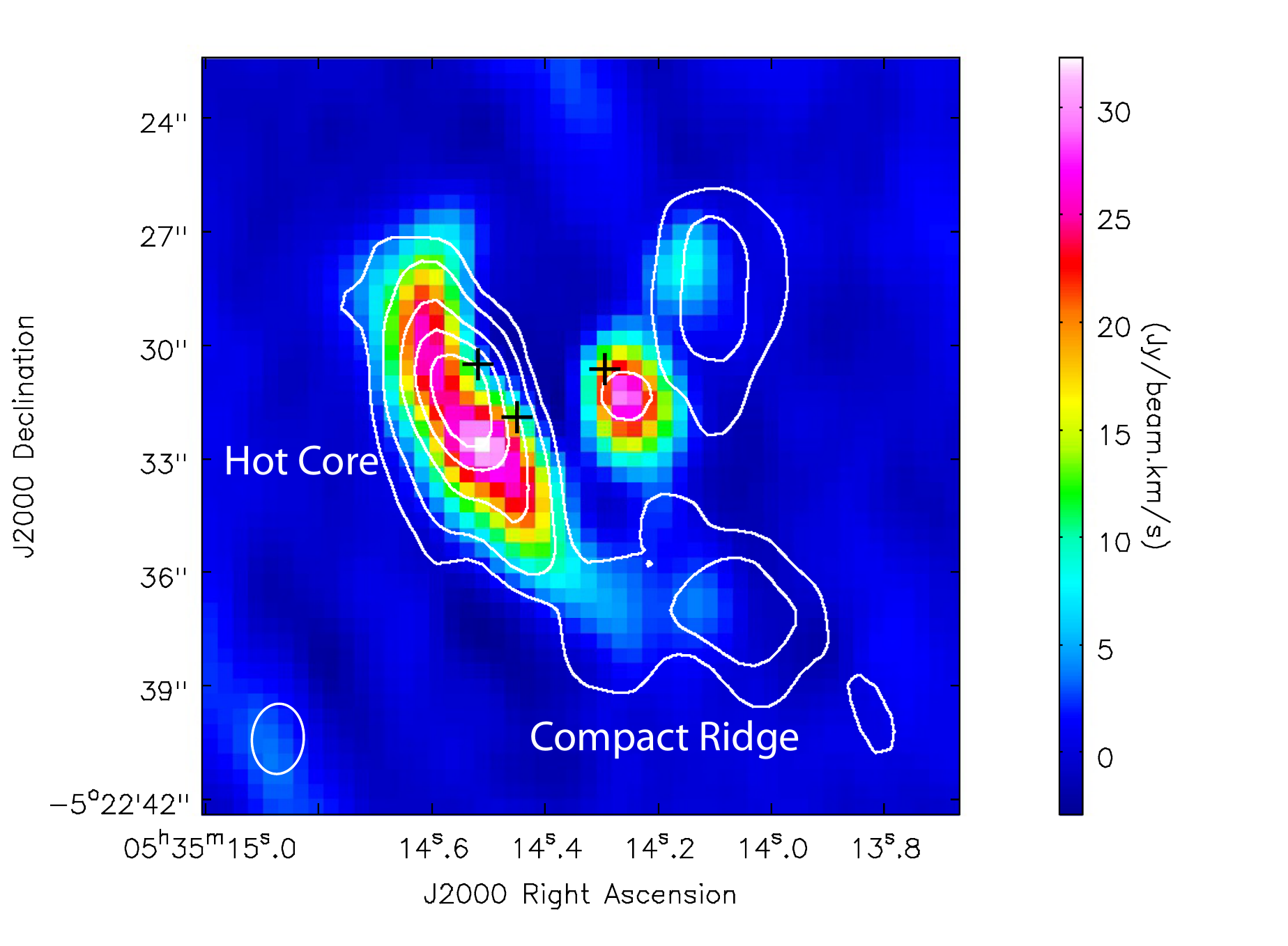}
\centering
\caption{ALMA moment 0 image of the $3_{22}$$^--3_{12}$$^+$ transition of NH$_2$D.  The color scale is integrated NH$_2$D flux over the velocity range 0.0--7.9 km s$^{-1}$.  The contours are the 230 GHz continuum emission, with contour levels (0.1, 0.2, 0.4, 0.6, 0.8) $\times$ 1.34 Jy beam$^{-1}$.  The three black crosses indicate the locations of sources I, SMA1, and IRc7 (from left to right).}
\end{figure}

The excitation of NH$_2$D transitions is likely to be complex, because of the high densities required to achieve local thermodynamic equilibrium (LTE) and the possible effects of radiative excitation.  For each detected transition, an upper-state column density can be derived using the following expression \citep{goldsmith99}:

\begin{equation}
N_u = \frac{1.67 \times 10^{17} W g_u}{\nu S_\textnormal{ij}\mu^2 \eta_{bf}}
\end{equation}

\noindent where $W$ is the integrated line flux (in K km s$^{-1}$), $\nu$ is the line frequency in MHz, $S_\textnormal{ij} \mu^2$ is the line strength in D$^2$, and $\eta_{bf}$ is the beam dilution factor.  This equation assumes that the line is optically thin, but is otherwise independent of the excitation of the molecule.  The $\Delta J = 1$, $\Delta K_c = 0$ transitions have the highest line strengths and so may not be optically thin.  Many of these lines also lie in bands 6 and 7 of HIFI (1425--1906 GHz) where attenuation from dust may be an important consideration.  There are several instances where multiple transitions with the same upper state are detected, which should yield the same $N_u$ from equation (1) if both transitions are optically thin.  As can be seen in Table 2, in the cases where a line in bands 1--5 has an upper state in common with a line in bands 6--7, the higher-frequency line yields a lower $N_u$ in the optically thin case than the lower-frequency transition by up to an order of magnitude, suggesting significant optical depth.  For the following analysis, we neglect the transitions detected in bands 6 and 7.  The remaining lines (28 of \emph{ortho} and 15 of \emph{para}) are assumed to be optically thin.  Finally, we note that the transition detected with ALMA shown in Figure 3 broadly follows the trend in a population diagram of the HIFI transitions (not shown), though this diagram has a large degree of scatter due to non-LTE excitation and possible optical depth effects.  Other transitions of NH$_2$D lie within the ALMA frequency bands (including a number of the lines in Table 2) and a comparison of single dish and interferometric observations can allow better constraint on the spatial distribution and excitation of deuterated ammonia than is possible here.

The total NH$_2$D column density was determined by summing the populations of all levels that can be determined using equation (1), which are related to the total NH$_2$D column density by

\begin{equation}
N_\textnormal{total} = f_c \sum N_\textnormal{observed}
\end{equation}

\noindent where $f_c$ is a correction factor to account for population in the levels not probed directly.  This factor is calculated on the basis of non-LTE statistical equilibrium calculations with the RADEX code \citep{vandertak07}.  Because rates for the collisional excitation of NH$_2$D by H$_2$ have only been calculated for low-energy levels and for colder temperatures than are found in the Orion Hot Core \citep{machin06}, we use the collisional rates available through the LAMDA database \citep{schoier05} for NH$_3$ \citep{danby88} to estimate the correction factor.  This treatment is only approximate because of the significant differences in the energy level structures and allowed selection rules for NH$_2$D and NH$_3$ as described above.  A kinetic temperature of 200 K and a H$_2$ density of $10^8$ cm$^{-3}$ are used for the correction factor calculation (from Table 1).  We also include a background continuum field which was based on observations with the Infrared Space Observatory \citep{vandishoeck98, lerate06} and HIFI \citep{bergin10}.  The observations with the larger ($80''$) ISO beam were scaled to match the HIFI survey at the shortest wavelength of HIFI (160 $\upmu$m), as is further described elsewhere \citep{crockett13a, neill13}.  A recent study of H$_2$S emission in Orion KL has found that reproducing the observed line fluxes, particularly on the highest energy transitions, requires a radiation field enhanced by a factor of 8 over this observed continuum field for $\lambda < 100 \upmu$m, implying a highly embedded continuum source \citep{crockett13a}.  Other studies of the emission of H$_2$O, HDO, and NH$_3$ \citep{hermsen88b, jacq90} have likewise suggested very high far-IR opacity in the Hot Core.  In this study, because we are using NH$_3$ as a proxy for NH$_2$D, it is not possible to draw firm conclusions about the excitation.  However, the correction factor for NH$_3$ is not strongly dependent on whether an enhanced far-IR radiation field is used.  The possibility of radiative pumping through vibrationally excited states, which has been proposed to be important elsewhere \citep{schilke92}, is not accounted for in the current analysis.  Correction factors of 2.0 are derived for \emph{ortho}-NH$_2$D, and 3.1 for \emph{para}.  We estimate an uncertainty of 25\% in these values, which dominates the uncertainty in the column density calculation.  This results in a total \emph{ortho}-NH$_2$D column density of $(6.0 \pm 1.6) \times 10^{15}$ cm$^{-2}$, and $(2.8 \pm 0.7) \times 10^{15}$ cm$^{-2}$ for para, or a total column density of $(8.8 \pm 1.7) \times 10^{15}$ cm$^{-2}$.  The derived \emph{ortho}:\emph{para} ratio is $2.1 \pm 0.7$, slightly lower but consistent with the statistical value of 3.  However, we note that \emph{ortho} transitions have higher optical depth than \emph{para}, which could lead to an underestimate of the \emph{ortho} column density; therefore, caution must be exercised in the interpretation of this result.

For NH$_3$, a total of 37 inversion transitions were observed between the studies of \cite{hermsen88b} and \cite{wilson93}.  Some systematic derivations from LTE conditions were observed in the level populations, with different column densities and rotational temperatures derived for each $K$ ladder, and from fits to the metastable levels and non-metastable levels.  The high-$J$ (10--14) inversion doublets measured by \cite{wilson93} suggest a thermal gradient, with a very hot ($T \sim 400$ K) component primarily responsible for this emission; the temperature map of hot NH$_3$ emission by \cite{goddi11} likewise reveals the presence of temperature gradients in this region.  With these caveats in mind, we nevertheless follow \cite{hermsen88b} in using a single-component rotation diagram to derive the total NH$_3$ column density of $(1.3 \pm 0.3) \times 10^{18}$ cm$^{-2}$ in a $6''$ source size, with a best-fit rotational temperature of $179 \pm 10$ K.  This yields a [NH$_2$D]/[NH$_3$] ratio of $(6.8 \pm 2.4) \times 10^{-3}$ in the Hot Core.

As mentioned above, we do not detect either a spike or plateau component in NH$_2$D transitions.  We derive upper limits for the [NH$_2$D]/[NH$_3$] ratio in these components using the linewidth parameters given in \citep{hermsen88b} (consistent with those in Table 1) and assuming LTE level populations at $T_\textnormal{rot} = 126$ K for the Plateau and 55 K for the spike as derived for NH$_3$.  This yields [NH$_2$D]/[NH$_3$] $<$ 0.01 in both of these components.

\subsection{Formaldehyde}

The lineshapes observed for the three detected isotopologues of formaldehyde in this study (H$_2$$^{12}$CO, H$_2$$^{13}$CO, and HDCO) indicate its presence in several environments within Orion KL, as seen in Figure 4.  Up to three Gaussian components are fit for H$_2$$^{12}$CO lines, the Plateau (shown in cyan in Figure 4), the Hot Core (in red), and the Compact Ridge (in green).  A total of 69 transitions of H$_2$$^{12}$CO are detected, ranging in lower-state energy from 70--980 K.  The highest-energy lines ($E_l > 550$ K) are well fit with only a Hot Core component.  For the H$_2$$^{13}$CO isotopologue, the Plateau component is not observed, and transitions are well fit with two components attributed to the Hot Core and Compact Ridge.  Again, the highest-energy transitions (in this case, those with $E_l > 200$ K) are emissive only from the Hot Core.

\begin{figure}
\figurenum{4}
\centering
\includegraphics[width=3.0in]{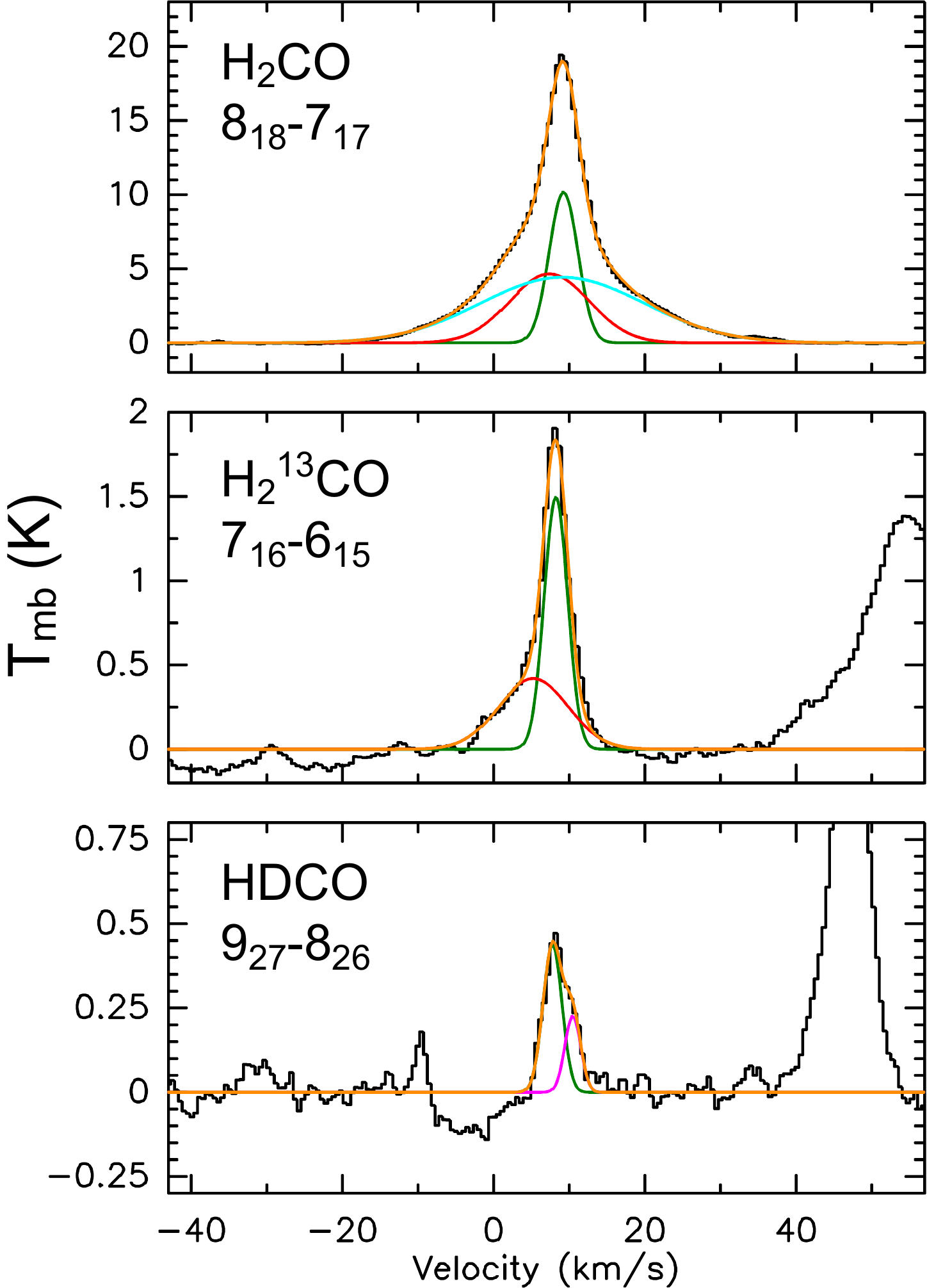}
\caption{Lineshapes of transitions of the three detected isotopologues of formaldehyde.  The individual fit Gaussian components are indicated by their color: green for the compact ridge, red for the hot core, cyan for the plateau, and magenta for the 10.6 km s$^{-1}$ component.  The total fit to each line is shown in orange.}
\end{figure}

The observed lines of HDCO show a different kinematic profile than the other two isotopologues.  Neither the Plateau nor the Hot Core components are detected, as seen in Figures 4 and 5.  Additionally, the HDCO line profile shows a wing to higher velocity ($v_\textnormal{LSR} \sim 10.4$ km s$^{-1}$) that is not clearly evident in the H$_2$$^{12}$CO or H$_2$$^{13}$CO lines, suggestive of a component with a higher HDCO/H$_2$CO ratio.  The HDCO lines are weak, so irregularities in the lineshape must be viewed with caution.  However, all of the unblended transitions of HDCO detected in the HIFI survey are shown in Figure 5, and the presence of this wing in nearly all of the transitions makes its detection more certain.  This asymmetry in the lineshape was also noted by \cite{loren85}.  Each of the detected HDCO lines was therefore fit with two Gaussian components, one with $v_\textnormal{LSR} \sim 7.6$ km s$^{-1}$ and $\Delta v \sim 3$ km s$^{-1}$ (the Compact Ridge), and the second with $v_\textnormal{LSR} \sim 10.4$ km s$^{-1}$ and $\Delta v \sim 2.5$ km s$^{-1}$.  In the $9_{37}-8_{36}$ line in Figure 5, only one component could be fit, as this is the weakest detected HDCO line, and the baseline is somewhat uncertain due to strong nearby lines.   In a number of these fits, the parameters of the two Gaussian components are highly correlated, and so the linewidths (and sometimes the central velocity) were fixed to the parameters given above.  \cite{loren85} proposed that the \mbox{10.4 km s$^{-1}$} component originates from a cloud to the north of the Compact Ridge, but complex organic molecules such as methyl formate (CH$_3$OCHO) have asymmetric lineshapes toward the Compact Ridge itself \citep{favre11}, so the spatial origin of this component is currently uncertain.  In Tables 3--5, the fit Gaussian parameters are presented for the transitions of formaldehyde isotopologues detected in this study.

\begin{figure}
\figurenum{5}
\centering
\includegraphics[width=5.0in]{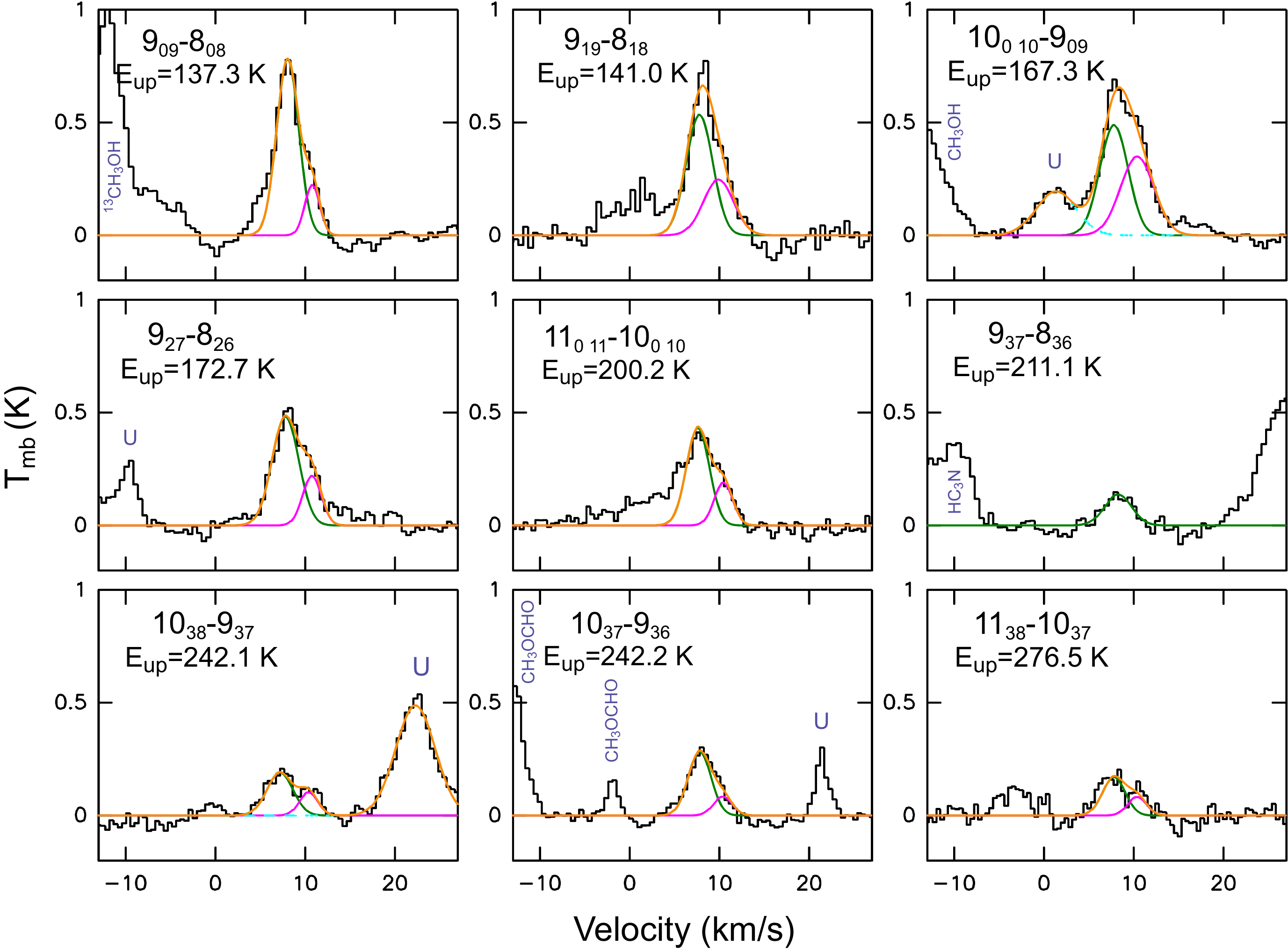}
\caption{All lines of HDCO detected in the HIFI survey.  The two fit components are shown in green (for 8.0 km s$^{-1}$, the nominal compact ridge) and magenta (for 10.4 km s$^{-1}$), and the total fit is shown in orange.  The weak $9_{37}-8_{36}$ transition is fit with only the 8 km s$^{-1}$ component.}
\end{figure}

Here we focus on determining the physical parameters of H$_2$CO in the Compact Ridge, leaving a detailed analysis of the Plateau and Hot Core emission aside since HDCO is not detected in these components.  The formaldehyde emission from the Compact Ridge was modeled using RADEX, assuming a uniform sphere geometry.  Collisional excitation rates for both \emph{para} and \emph{ortho}-H$_2$CO with H$_2$, available in the LAMDA database, are available for temperatures up to 300 K and involving rotational energy levels up to $E = 300$ K.  These rates come from the calculations of \cite{green91} for the interaction of H$_2$CO with He, and are scaled to estimate rates with H$_2$ as the collider.  A recent study calculating the excitation of H$_2$CO with H$_2$ found that the rates of \cite{green91} may have significant and systematic errors and that the rates may differ for collisions with \emph{ortho} and \emph{para}-H$_2$ \citep{troscompt09}.  However, rates were calculated in that study only for lower energy states ($J \le 5$) than are studied with HIFI, and only up to $T = 100$ K, so for the present study we used the He-scaled rates of \cite{green91}.  Because transitions of H$_2$CO are observed in the Compact Ridge from states with energies higher than have been calculated directly, the collisional rates of \cite{green91} were extrapolated to higher energies using the artificial neural network code of \cite{neufeld10}.  For HDCO, due to the lack of symmetry, the selection rules are more complex, as both $a$-type and $b$-type transitions are allowed.

Transitions with $J' \ge 7$ are observed within the HIFI bandwidth for each isotopologue.  This leads to a degeneracy between the formaldehyde column density and the excitation conditions; that is, population can be ``hidden" in the levels that are not probed by HIFI if the H$_2$ density is reduced (making the excitation more subthermal) and the H$_2$CO column density is increased.  To break this degeneracy, we also include in our models the three $J = 2-1$ transitions of both H$_2$CO and H$_2$$^{13}$CO measured by the Turner 2 mm (130--170 GHz) survey program with the 12 m NRAO telescope \citep{remijan08}.\footnote{This survey is publicly available at \url{http://www.cv.nrao.edu/Turner2mmLineSurvey/.}}  This survey was chosen due to the similarity in the beam size of this survey ($44''$ at 140 GHz) with that of the HIFI survey ($42''$ at 500 GHz).  The pointing position is slightly different between the two surveys by about $5''$---the Turner survey was centered on the Hot Core, while the HIFI survey is centered in between the Hot Core and Compact Ridge---but with the large beam size of the Turner survey compared to the regions of compact emission (see Fig. 3), this difference is probably not significant.  Finally, as the 2 mm transitions can be excited at lower temperatures and densities than those observed in HIFI, the potential contribution of the Extended Ridge to the $J = 2-1$ lines should not be neglected.  The released data product is on an antenna temperature scale, which we convert to main beam tempratures using a main beam efficiency of 0.75 (A. Remijan, private communication).  The transitions used from this survey and our lineshape fits, performed the same way as for the HIFI transitions, are presented in Figure 6.

\begin{figure}
\figurenum{6}
\centering
\includegraphics[width=5.0in]{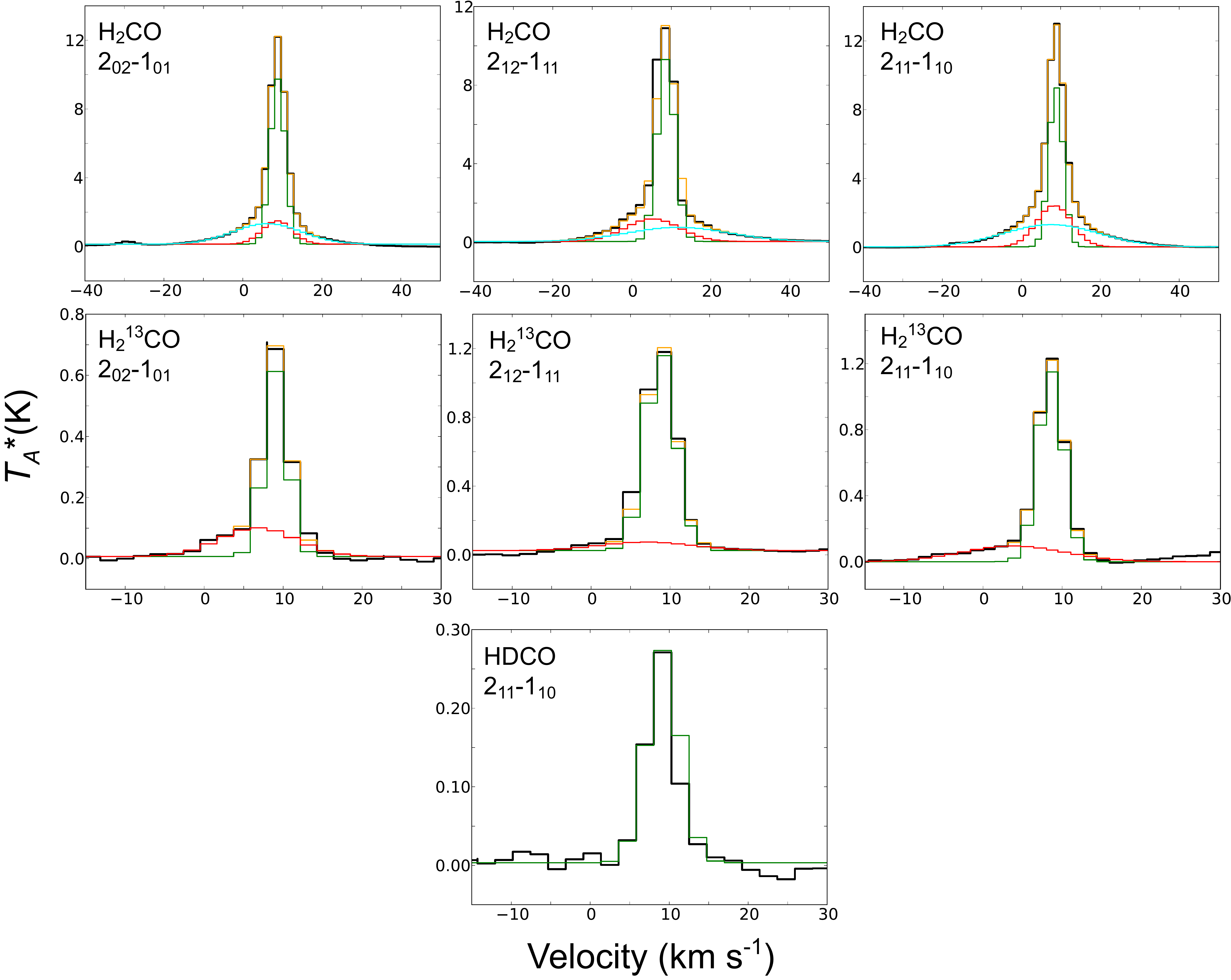}
\caption{Transitions of H$_2$$^{12}$CO, H$_2$$^{13}$CO, and HDCO detected in the Turner 2 mm survey.  In each panel, the orange curve represents the full fit to the data, while the other curves represent the fits to the Plateau (in cyan), Hot Core (in red), and Compact Ridge (in green).}
\end{figure}

A total of 12 transitions of H$_2$$^{13}$CO have an identified compact ridge component in the HIFI survey.  A rotation diagram analysis \citep{goldsmith99}, assuming all lines are optically thin, yields a rotation temperature of $40.0 \pm 2.3$ K.  This could indicate either dense ($\ge 10^7$ cm$^{-3}$), thermalized gas at a kinetic temperature close to 40 K, or subthermal gas that is warmer than this temperature but with a lower H$_2$ density.  We favor the latter interpretation.  The corresponding lines of H$_2$CO are optically thick, and so the observed line brightness temperature is given by $\eta_{bf} J(T_\textnormal{ex})$, the source function of the kinetic temperature in the gas multiplied by the excitation temperature.  Under the conditions of the Compact Ridge, the derived excitation temperature is a lower limit to the true gas kinetic temperature.  Assuming a source size of $15''$, consistent with interferometric observations from \cite{mangum93}, we derive $T_\textnormal{ex} = 70-100$ K for the most optically thick lines of H$_2$CO, reasonably consistent with the canonical temperature range for the Compact Ridge given in Table 1.  If a smaller source size of $10''$ is adopted, an excitation temperature of $120-190$ K is calculated, higher than typically derived for this region.  We therefore adopt $15''$ as the source size of the formaldehyde emission for the following analysis.

All detected transitions of H$_2$CO and H$_2$$^{13}$CO in the Compact Ridge can be fit well with a two component model, as seen in Figure 7.  For this model we assume a $^{12}$C/$^{13}$C ratio of 60 \citep{persson07}, and an \emph{ortho}:\emph{para} ratio of 3:1 (the high-temperature limit).  The physical properties of these components are given in Table 6.  One component, which is attributed to warm, moderate-density ($n$(H$_2$)= $1.8 \times 10^{6}$ cm$^{-3}$) gas, is responsible for most of the flux to the low-energy lines, including the H$_2$$^{13}$CO lines.  The second component is warmer and about an order of magnitude denser, and is attributed to a clump or multiple clumps of gas in the Compact Ridge:  interferometric observations of the Compact Ridge region reveal several such clumps, each a few arcseconds in size \citep{favre11, peng12}.  In Figure 7, the plotted fluxes are the sum of the fluxes predicted by the two components individually.  We assume a 25\% uncertainty in the H$_2$CO column density in both components, dominated by uncertainty in the excitation.  It can be seen that the low-energy ($E_l < 50$ K) H$_2$$^{12}$CO lines (the three lines from the Turner survey) are underpredicted by this model by about a third, which could be due to emission from the Extended Ridge.

\begin{figure}
\figurenum{7}
\centering
\includegraphics[width=6.0in]{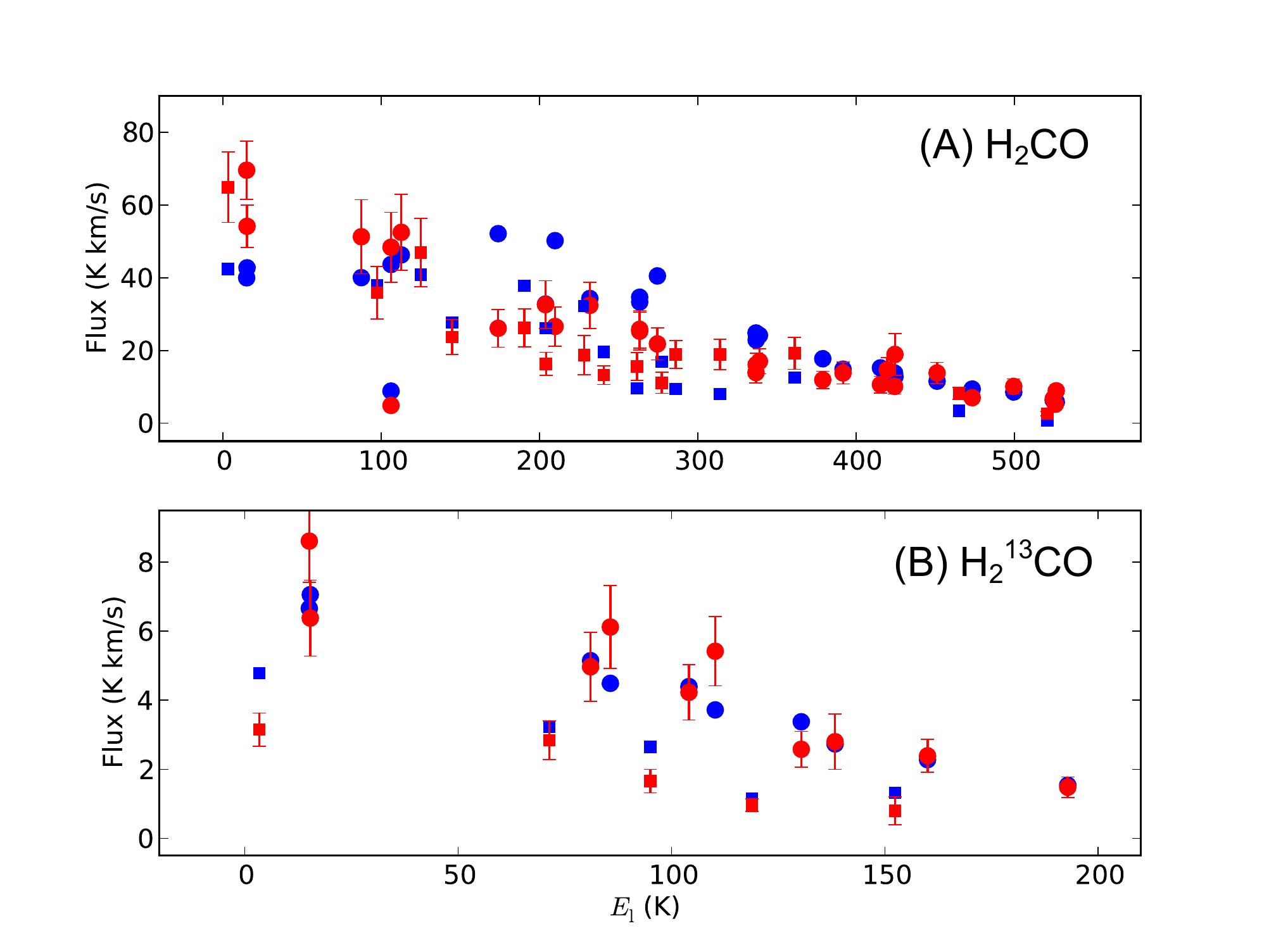}
\caption{Two-component model to transitions of H$_2$CO and H$_2$$^{13}$CO in the Compact Ridge, assuming a $^{12}$C/$^{13}$C ratio of 60.  The excitation parameters used for the two components are given in Table 6.  The red points, with error bars, indicate the measured integrated flux from the HIFI survey, while the blue points indicate the fluxes calculated by the model (derived by summing together the fluxes calculated for two components).  Squares indicate lines of \emph{para} formaldehyde, while circles indicate lines of \emph{ortho}.}
\end{figure}

\begin{deluxetable}{c c c c c}
\tablenum{6}
\tablewidth{0pt}
\tablecaption{Fit parameters for the two-component non-LTE model to H$_2$$^{12}$CO and H$_2$$^{13}$CO emission in the Compact Ridge.}
\tablehead{ Component & $\theta_s$ & $N$(H$_2$CO) & $T$ & $n$(H$_2$) \\
                                       &     $('')$      &    (cm$^{-2}$)   & (K)   &     (cm$^{-3}$)}
\startdata
1 & 15 & $5.0 \times 10^{16}$ & 125 & $1.8 \times 10^6$ \\
2 &  5  & $8.0 \times 10^{15}$ & 175 & $2.0 \times 10^7$ \\
\enddata
\end{deluxetable}

As shown in Figure 5, a total of nine transitions of HDCO are detected.  Figure 8 shows population diagrams of H$_2$$^{13}$CO and HDCO transitions.  The top panel of this figure shows the transitions of H$_2$$^{13}$CO:  the observed values are in red, while the non-LTE model in Figure 7 is in blue.  The red line indicates the best linear fit, yielding $T_\textnormal{rot} = 40.0 \pm 2.3$ K as noted above.  In this fit, we neglect the three transitions observed from the 2 mm survey because of the possible contribution from the Extended Ridge (shown as unfilled points).  The lower panel shows the detected lines of HDCO; the dashed lines indicate linear fits to the two velocity components.  Both fits yield significantly higher excitation temperatures than for H$_2$$^{13}$CO ($T_\textnormal{rot} = 67$ and 63 K for the 8 and 10.4 km s$^{-1}$ components, respsectively).  This could be attributed to a few different causes.  First, there may be a gradient in the D/H fractionation in the Compact Ridge, and the HDCO may emit from warmer and/or denser gas than H$_2$$^{13}$CO.  Alternatively, it could be an excitation effect.  Collisional rates for HDCO are not known, but the lower molecular symmetry of HDCO means that $\Delta K_a = 1, 3, 5...$ collisional and radiative transitions are allowed, unlike in H$_2$CO and H$_2$$^{13}$CO.  The radiative $\Delta K_a = 1$ transitions (\emph{b}-type transitions) are very weak (line strengths roughly 2 orders of magnitude lower than \emph{a}-types) and so likely contribute little to the relaxation.  However, if collisional \emph{b}-type transitions have rates that are competitive with \emph{a}-types, this would produce more pathways for collisional excitation, and so would drive the excitation more toward $T_\textnormal{kin}$ than for H$_2$$^{13}$CO.  Also, the $J \rightarrow J+1$ rotational spacing ($\sim B + C$) in HDCO is 10\% lower than in H$_2$$^{13}$CO due to its higher moment of inertia along the \emph{a} principal axis.  While HDCO has the same total dipole moment as H$_2$$^{13}$CO, the Einstein A coefficient of an HDCO transition is about 25\% less than the corresponding transition of H$_2$$^{13}$CO due to the factor of $\nu^3$.  This would have the same effect of bringing the level populations in HDCO closer to $T_\textnormal{kin}$ under the same conditions.  Interferometric observations will be needed to determine the relative spatial distributions of H$_2$CO and HDCO, while collisional rates that include the states detected by HIFI will be needed to judge the impact of the excitation differences postulated here.

\begin{figure}
\figurenum{8}
\centering
\includegraphics[width=6.0in]{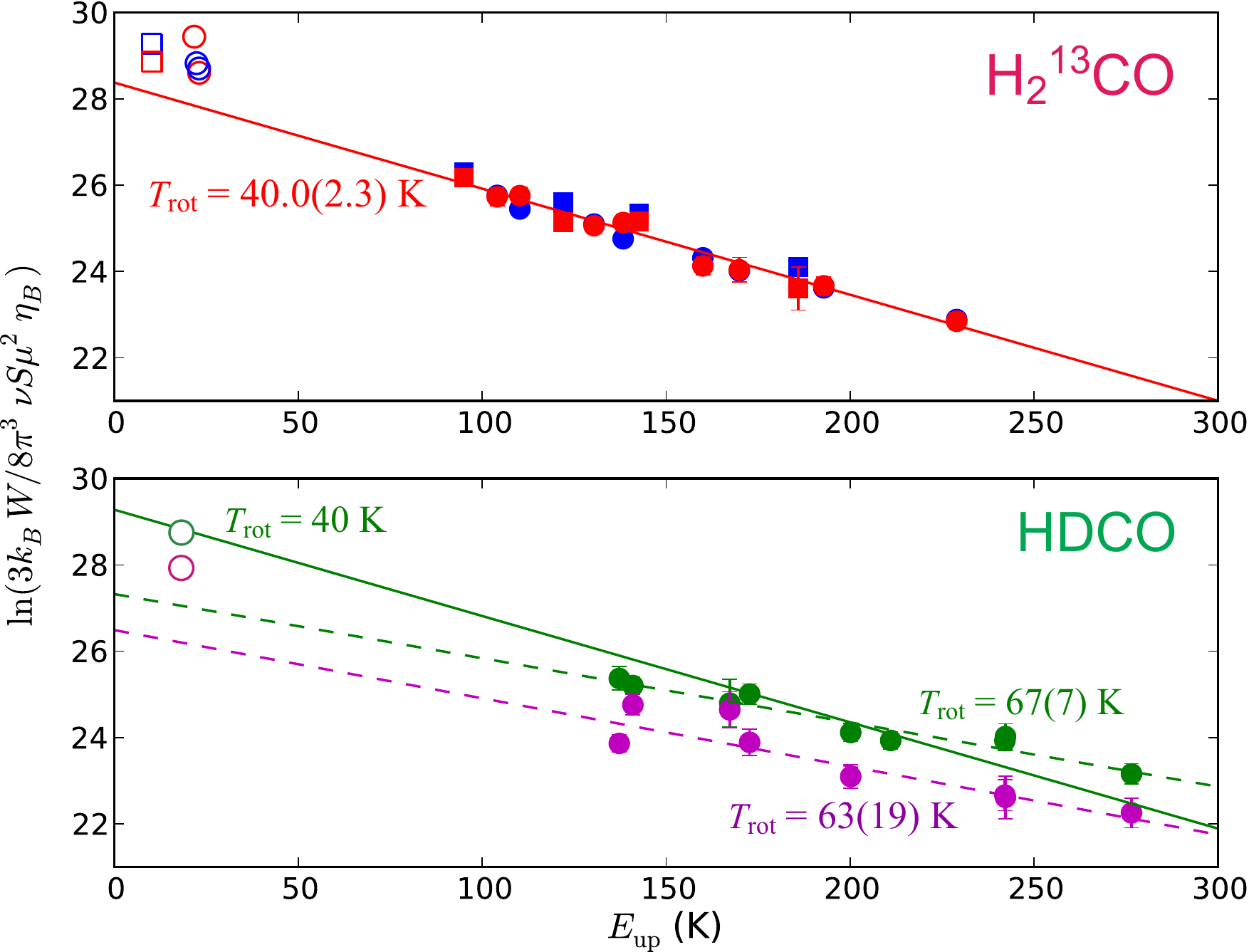}
\caption{Population diagrams of H$_2$$^{13}$CO and HDCO transitions.  In the top panel, the red points are the observed lines from HIFI and the Turner 2 mm survey, and the blue points are the calculated flux from the RADEX model presented in Figure 7.  The red line is a linear fit to the observed (red) data points.  In the bottom panel, the green points indicate the observed fit fluxes of HDCO in the compact ridge ($\sim 8.0$ km s$^{-1}$) component, while the magenta points indicate the observed fluxes in the 10.6 km s$^{-1}$ component.  The open points indicate the transitions from the Turner survey; for the HDCO transition, we assume that 2/3 of the flux is due to the 8.0 km s$^{-1}$ component and 1/3 is due to the 10.6 km s$^{-1}$ component (see the text for further discussion).  In cases where error bars are not visible, they are roughly the size of the data points.  A $15''$ source size is assumed for both components.}
\end{figure}

If we take the $\sim 65$ K excitation temperatures derived from the fit to the HDCO components, assuming optically thin conditions, we derive a column density of $(1.5 \pm 0.6) \times 10^{14}$ cm$^{-2}$ for the 8 km s$^{-1}$ component, and $(6.0 \pm 5.4) \times 10^{13}$ cm$^{-2}$ for the 10.6 km s$^{-1}$ component.  If instead we neglect the possible excitation differences between HDCO and H$_2$$^{13}$CO and assume the same excitation temperature as for H$_2$$^{13}$CO (40 K), as indicated by the solid green line in the lower panel of Figure 8, we get an HDCO column density of $(5.1 \pm 1.3) \times 10^{14}$ cm$^{-2}$ for HDCO in the 8 km s$^{-1}$ component.  The discrepancy between these two values is a factor of 3.4 and suggests significant uncertainty in the excitation of HDCO.  There is one transition of HDCO in the Turner NRAO survey that can help to distinguish between these two scenarios, the $2_{11}-1_{10}$ transition at 134284.83 MHz.  The other two $J = 2-1$ transitions of HDCO are outside the frequency coverage of this survey.  Due to the relatively low spectral resolution of the 2 mm survey (0.8 MHz, or 1.8 km s$^{-1}$), attributing the flux to the two narrow velocity components detected in HIFI is not possible.  The line is well fit with a single Gaussian with $v_\textnormal{LSR} = 8.2$ km s$^{-1}$ and $\Delta v = 5.0$ km s$^{-1}$.  For illustrative purposes, we attribute 2/3 of the flux to the 8 km s$^{-1}$ component and 1/3 to the 10.6 km s$^{-1}$ component, in keeping with most of the HIFI lines (see Figure 5), and plot the resulting points as open circles in Figure 8.  These data points are not used in any fits.  The flux assigned to the 8 km s$^{-1}$ component lies well above the 67 K fit line, and in good agreement with the 40 K line.  However, as for H$_2$$^{13}$CO, the possible contribution of flux from the Extended Ridge should be considered.  Therefore, we adopt the average of the HDCO column densities derived by the two methods, yielding a final value of $3.3 \times 10^{14}$ cm$^{-2}$ for the 8 km s$^{-1}$ Compact Ridge component, with an assumed uncertainty of a factor of 2 due to the uncertainty in the excitation.  Comparing this value with the H$_2$CO abundance from the RADEX model results in a [HDCO]/[H$_2$CO] ratio of $0.066_{-0.037}^{+0.068}$.

As seen in Figure 5, HDCO is not detected in the hot core or plateau components.  In order to establish upper limits in these components, we employ the H$_2$CO LTE models from the Orion KL full-band analysis (Crockett et al. 2013b, in preparation).  In the Hot Core, due to the high H$_2$ densities ($\ge 10^7$ cm$^{-3}$), the LTE approximation is likely valid.  H$_2$CO is likely very optically thick, but the hot core component is detected in 26 lines of H$_2$$^{13}$CO up to $E_\textnormal{up} = 500$ K.  The best fit LTE model has $\theta_s = 10''$, $T_{ex} = 120$ K, $v_\textnormal{LSR} = 6.5$ km s$^{-1}$, $\Delta v = 13$ km s$^{-1}$, and $N$(H$_2$$^{13}$CO) = $6.0 \times 10^{14}$ cm$^{-2}$.  Applying the same temperature and kinematic parameters for HDCO, we find the upper limit on $N$(HDCO) in the hot core to be $1.8 \times 10^{14}$ cm$^{-2}$, which implies [HDCO]/[H$_2$CO] $<$ 0.005.  In the plateau, the best fit parameters for H$_2$CO are $\theta_s = 30''$, $T_\textnormal{ex} = 115$ K, $v_\textnormal{LSR} = 8$ km s$^{-1}$, $\Delta v = 22$ km s$^{-1}$, and $N$(H$_2$CO) = $1.3 \times 10^{15}$ cm$^{-2}$.  Using the same parameters except for the HDCO column density, we find an upper limit to $N$(HDCO) of $2.5 \times 10^{13}$ cm$^{-2}$, or [HDCO]/[H$_2$CO] $<$ 0.019.

We also do not detect D$_2$CO in this study, although this species has been detected previously in Orion KL by \cite{turner90}.  To set an upper limit on its abundance in the Compact Ridge, we use an LTE model, assuming $T_\textnormal{rot} = 60$ K and $\theta_s = 15''$, and the kinematic parameters of the HDCO transitions.  This yields [D$_2$CO]/[HDCO] $\le 0.1$.  \cite{turner90} detected three transitions of D$_2$CO in the submillimeter and derived a [D$_2$CO]/[HDCO] ratio of 0.02, consistent with our finding here.

\subsection{Methanol}

Methanol and its isotoplogues have more detected transitions in the HIFI survey than any other molecule (Crockett et al. 2013b, in preparation).  A study of CH$_3$OH and $^{13}$CH$_3$OH in this survey has been published \citep{wang11}, focusing on two $Q$-branches, which have a large number of transitions spanning a wide energy range in a narrow frequency space.  This analysis found that the bands were best reproduced using an externally heated thermal gradient; the idea that the compact ridge is externally heated was initially proposed by \cite{blake87}.  The two singly deuterated forms of methanol, CH$_2$DOH and CH$_3$OD, have also both been detected previously in Orion KL.  The first analysis of both species, by \cite{jacq93}, found a column density [CH$_2$DOH]/[CH$_3$OD] = 1.1--1.5, below the statistical ratio of 3 that would be expected if each of the four hydrogen atoms in methanol were deuterated in equal proportion.  This ratio has also been examined recently by \cite{peng12} using the Plateau de Bure interferometer; in this study, the CH$_2$DOH/CH$_3$OD ratio was found to be less than 1 over the compact ridge region.

We use the unprecedented frequency coverage of the HIFI survey, and the large number of transitions of each isotopologue in the bandwidth, along with new comprehensive spectroscopic analyses of CH$_2$DOH and CH$_3$OD, to get the best constraint on the abundances of these species.  As collisional rates are only available for CH$_3$OH for relatively low-lying levels (up to $J = 15$) \citep{rabli10}, we use LTE modeling in XCLASS to derive the column densities in each of these species.  \cite{wang11} found that a thermal gradient was needed to describe the emission of the $^{12}$CH$_3$OH band from the compact ridge.  However, the $^{13}$CH$_3$OH band in that study was well modeled by a single LTE component due to the nondetection of high energy ($E_u > 400$ K) transitions that reveal the gradient most clearly.   In the models used here, we have accounted for all transitions in the HIFI survey bandwidth, so the parameters vary somewhat from those presented by \cite{wang11}.  Because of its high optical depth, we neglect $^{12}$CH$_3$OH for this analysis, and derive the methanol column density using  $^{13}$CH$_3$OH.

The $^{13}$CH$_3$OH line profiles are well modeled by two velocity components.  The spatial origin of these components is investigated in Figure 9, compared to the HDO emission as presented in \cite{neill13}.  In the top two panels, the profiles of a HIFI transition and an ALMA transition with similar excitation energy are shown.  The ALMA line profile in panel B is smoothed to a $20''$ beam.  Both transitions are well modeled by two Gaussian components, one narrow ($v_\textnormal{LSR} \sim 8$ km s$^{-1}$, $\Delta v$ = 2.5 km s$^{-1}$) and one broader ($v_\textnormal{LSR} \sim 7.5$ km s$^{-1}$, $\Delta v \sim 7$ km s$^{-1}$).  In panel C, showing the $^{13}$CH$_3$OH emission integrated from 7--9 km s$^{-1}$, where both velocity components contribute, there are a number of emissive clumps, with the three strongest peaks in the Compact Ridge region, offset to the southwest from the Hot Core emission (seen as the region of strongest millimeter continuum emission, in white contours in Figure 9).  The morphology of methanol emission in this region has been laid out in more detail by \cite{peng12}.  Panel D shows the $^{13}$CH$_3$OH emission integrated from 3--5 km s$^{-1}$, where only the broader velocity component contributes to the lineshape.  The peak of the emission in this velocity range is in the Hot Core region, coincident with the region of brightest HDO emission in Orion KL.  The velocity parameters for the broader component are also in very good agreement with the parameters for the HD$^{18}$O transitions detected by HIFI \citep{neill13}, and with the transitions of NH$_2$D as described in \S 3.2.  We adopt source sizes of $10''$ for both the Compact Ridge and Hot Core.

\begin{figure}
\figurenum{9}
\includegraphics[width=6.0in]{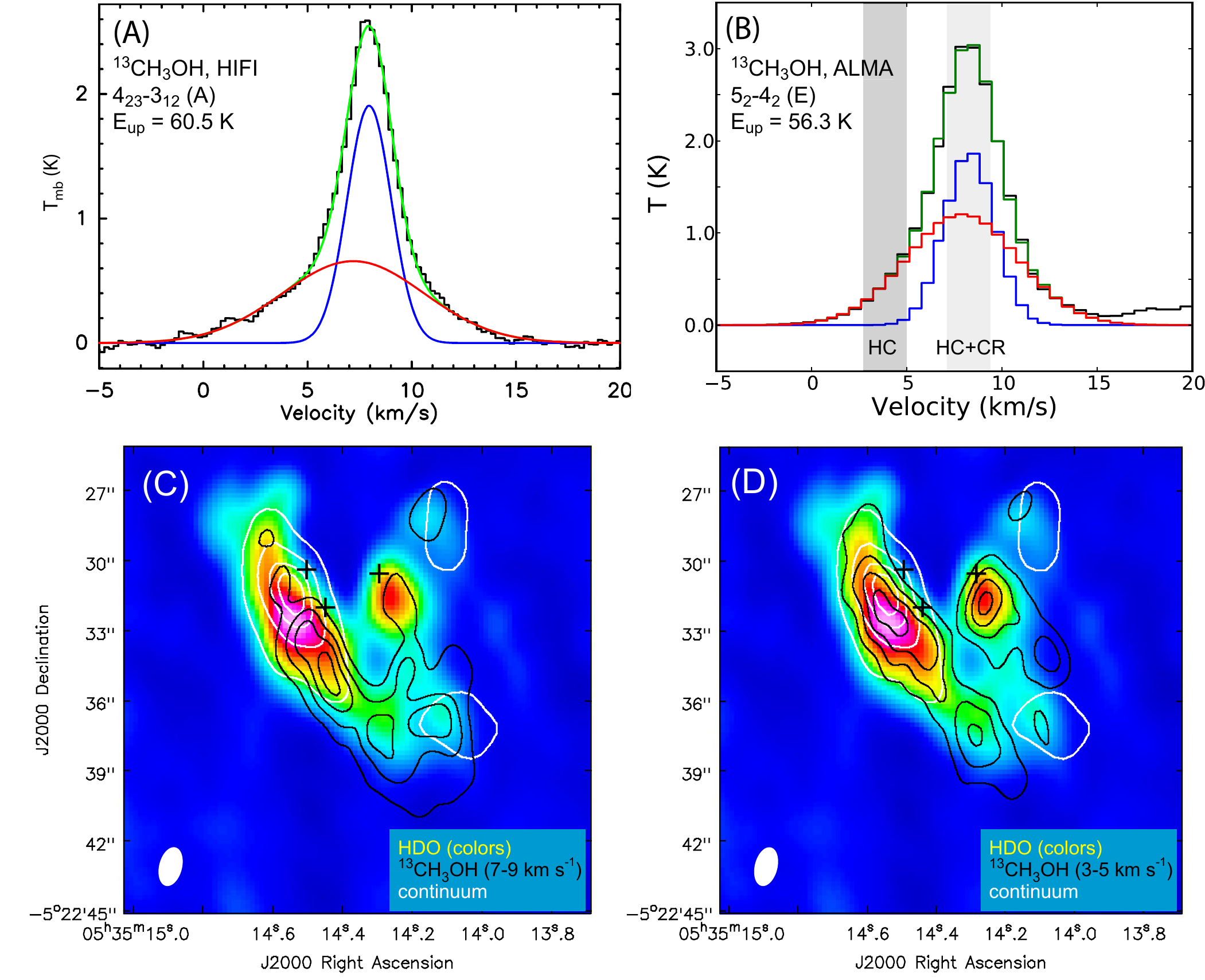}
\caption{Spatial distribution of $^{13}$CH$_3$OH in Orion KL.  Panel A: HIFI spectrum of the $4_{23}-3_{12}$ ($A$) transition, with a two-component fit:  the Compact Ridge component in blue, the Hot Core component in red, and the cumulative fit in green.  Panel B: ALMA spectrum of the $5_2-4_2$ ($E$) transition.  For this spectrum the spatial resolution has been smoothed to $20''$ to include the emission from all of the compact components within Orion KL.  The fit components are as in Panel A.  Panel C: $^{13}$CH$_3$OH emission integrated from 7--9 km s$^{-1}$ (black contours, with levels (0.3, 0.5, 0.7, 0.9) $\times$ 9.1 Jy beam$^{-1}$ km s$^{-1}$); HDO integrated emission from the $3_{12}-2_{21}$ transition at 225896.7 MHz (color scale, from \cite{neill13}); 230 GHz continuum emission (white contours, with levels (0.2, 0.5, 0.75, 0.9) $\times$ 1.34 Jy beam$^{-1}$). Panel D: $^{13}$CH$_3$OH emission integrated from 3--5 km s$^{-1}$ (black contours, with levels (0.3, 0.5, 0.7, 0.9) $\times$ 2.5 Jy beam $^{-1}$ km s$^{-1}$).  HDO and continuum emission are as in panel C.}
\end{figure}

Figures showing all of the unblended transitions of each species are presented as Figure Sets 10--12.  The parameters of the models used for these figures are presented in Table 7.  The deuterated isotopologues show only weak evidence for the broader velocity component (the Hot Core), so the column densities in this component are viewed as an upper limit.  The source size, temperature, and linewidth parameters are fixed to those of $^{13}$CH$_3$OH.  Dust opacity is also included in the model as part of the fullband analysis, using the H$_2$ column densities from Table 1 (Crockett et al. 2013b, in preparation), but is a minor effect in bands 1--5 ($\tau_d =  0.17$ for the Compact Ridge, and 0.14 for the Hot Core, at 1280 GHz).  The column densities derived here correspond to methanol abundances relative to H$_2$ of $1.5 \times 10^{-6}$ in the Compact Ridge and $2.1 \times 10^{-6}$ in the Hot Core, using H$_2$ column densities from Table 1.

\begin{deluxetable}{c c c c c c c c c c c}
\tablenum{7}
\tabletypesize{\footnotesize}
\tablewidth{0pt}
\tablecaption{Fit LTE parameters for isotopologues of methanol.}
\tablehead{                      & \multicolumn{5}{c}{Compact Ridge} & \multicolumn{5}{c}{Hot Core} \\
		Isotopologue & $\theta_s$ & $T_{rot}$ & $N$ & $v_{LSR}$ & $\Delta v$ & $\theta_s$ & $T_{rot}$ & $N$ & $v_{LSR}$ & $\Delta v$ \\
				    & $('')$          & (K)           & (cm$^{-2}$) & (km s$^{-1}$) & (km s$^{-1}$) & $('')$          & (K)           & (cm$^{-2}$) & (km s$^{-1}$) & (km s$^{-1}$)}
\startdata
$^{13}$CH$_3$OH & 10 & 140 & $1.0 \times 10^{16}$ & 2.5 & 8.0 & 10 & 128 & $1.1 \times 10^{16}$ & 6.5 & 7.5 \\
CH$_2$DOH & 10 & 140 & $3.5 \times 10^{15}$ & 2.5 & 8.0 & 10 & 128 & $\le 2.8 \times 10^{15}$ & 6.5 & 7.5 \\
CH$_3$OD & 10 & 140 & $3.0 \times 10^{15}$ & 2.5 & 8.0 & 10 & 128 & $\le 1.2 \times 10^{15}$ & 6.5 & 7.5 \\
\enddata
\end{deluxetable}

The large number of detected transitions for each isotopologue, as presented in the Appendix, serves as definitive evidence for their detection in the HIFI survey.  However, for a significant number of transitions the intensity of the LTE model overpredicts the line intensity.  For CH$_2$DOH, the spectroscopy is complicated by the reduction in symmetry; the line strengths used in this analysis come from the recent work of \cite{pearson12}, and are found in the current JPL catalog.  The \emph{a}-type intensities are considered to be more reliable than for \emph{b}- and \emph{c}-type transitions, so we have separated the \emph{a}-type CH$_2$DOH lines (110 transitions) in the online-only figures from the \emph{b}- and \emph{c}-type transitions.  The model fits the \emph{a}-type transitions reasonably well, with the exception of a few transitions that are clearly missing (no blends, and a noise level far below the intensity predicted by the model).  These transitions have similar parameters (frequency, lower-state energy, and line strength) to other transitions that are clearly detected, so this is likely to be due to catalog errors (either in frequency or intensity) for these lines.

While the catalog intensities for the other methanol isotopologues are likely more reliable than for CH$_2$DOH, there may still be issues.  For CH$_3$OH, a recent laboratory study has found errors in the line strengths of up to 50\% for some transitions \citep{fortman12a}.  Methanol line lists are calculated using molecular parameters from global fits combined with dipole moment information from the literature.  The quality of line intensities relies to a large extent on our knowledge of the dipole moment information available for the molecule. The line lists are calculated using experimentally determined permanent dipole moment values, except in the case of CH$_3$OH where ab initio torsional dependent dipole moment information was used \citep{mekhtiev99}.  The observed discrepancies call for an investigation of the torsional dependence of the dipole moment.  A second source of discrepancies exists for unresolved or barely resolved $A+/-$ transitions, where the current Hamiltonian model tends to either over- or underpredict the line intensities unless these lines are treated with special care.

Some transitions of $^{13}$CH$_3$OH and CH$_3$OD are overpredicted by the HIFI model (see the online-only figures).   Most of the overpredicted transitions lie at higher frequencies (above 800 GHz).  These also tend to be the lines with the highest line strengths and have the highest predicted optical depth in our LTE models.  (Many of these lines also happen to consist of two transitions with the same $J$ and $K_a$ separated by less than the linewidth, further increasing the opacity by a factor of two.)  This suggests that the model may be underpredicting the optical depth in these lines.  Increasing the optical depth of these lines with the single-component compact ridge model we use here would require substantially decreasing the source size, which we do not do here because it would be inconsistent with interferometric observations from Figure 9 and \cite{peng12}.  Higher extinction from dust may also reduce the intensity in the higher-frequency lines.  Finally, this could be explained by a thermal gradient as suggested by \cite{wang11}.  Because of the large number of transitions and isotopologues involved, we have not pursued more sophisticated modeling of the source for this study.  Most of the transitions we observe are consistent with the LTE model given in Table 7, particularly the weaker lines that are most likely to be optically thin in all components.  We assume 20\% uncertainties in each of the column densities.

From these models, we derive a ratio between the two deuterated forms of methanol, [CH$_2$DOH]/[CH$_3$OD], of $1.2 \pm 0.3$, consistent with the value of 1.1--1.5 that was found by \cite{jacq93}.  \cite{peng12} derived abundances for the three forms of methanol using Plateau de Bure observations.  They find lower overall levels of deuteration than this study: in the main methanol clumps in the Compact Ridge, they find [CH$_2$DOH]/[CH$_3$OH] $\sim 10^{-3}$, whereas we derive a value of 0.0058 for the compact ridge as a whole (assuming $^{12}$C/$^{13}$C = 60).  Additionally, while we find [CH$_2$DOH]/[CH$_3$OD] = $1.2 \pm 0.3$, they find this ratio to be less than 1 in each of the main clumps.  This ratio is derived using single transitions of CH$_2$DOH and CH$_3$OD with similar upper-state energies.  The CH$_2$DOH line used ($3_{13}-2_{02}$ of the $o_1$ torsional level) is a \emph{b}-type transition, and therefore, as noted above, its line strength may not be reliable.  They used $S_{ij} \mu^2 = 2.28$ D$^2$ for this transition, while the JPL catalog entry has a value of 1.69 D$^2$; adopting this latter value increases the CH$_2$DOH column density by 35\%, bringing the [CH$_2$DOH]/[CH$_3$OD] they find to approximately 1 (instead of 0.7---0.8 in their results), in close agreement with the value we derive here.

\section{Discussion}

\begin{deluxetable}{c c c}
\tablenum{8}
\tablewidth{0pt}
\tablecaption{D/H ratios in Orion KL.}
\tablehead{Ratio & Hot Core & Compact Ridge}
\startdata
[HDO]/[H$_2$O] & $3.0_{-1.7}^{+3.1} \times 10^{-3}$ & $3.8_{-2.5}^{+3.6} \times 10^{-3}$ \\ [0ex]
[NH$_2$D]/[NH$_3$] & $6.8_{-2.4}^{+2.4} \times 10^{-3}$ & $\le 10^{-2}$ \\ [0ex]
[HDCO]/[H$_2$CO] & $\le 5.0 \times 10^{-3}$ & $6.6_{-3.7}^{+6.8} \times 10^{-3}$ \\ [0ex]
[CH$_2$DOH]/[CH$_3$OD] & $\le 4.2 \times 10^{-3}$  & $5.8_{-1.2}^{+1.2} \times 10^{-3}$ \\ [0ex]
[CH$_3$OD]/[CH$_3$OH] & $\le 1.8 \times 10^{-3}$ & $5.0_{-1.0}^{+1.0} \times 10^{-3}$ 
\enddata
\end{deluxetable}

The D/H ratios derived in this study are summarized in Table 8.  In this table the HDO/H$_2$O ratios derived for the Hot Core and Compact Ridge from \cite{neill13} are also included.  These results differ somewhat from previous ratios derived for these molecules, but as noted above, due to the large number of transitions detected per species, we believe that these numbers are the most reliable determinations to date of these ratios.  It can be seen that all of the molecules detected here have similar D/H ratios (or upper limits), between (2---8) $\times$ $10^{-3}$, in the two components.  In particular, water has approximately the same deuterium fractionation as the other species.  This differs from previous observations toward lower-mass protostars, as well as in our own Solar System.  The low-mass protostar for which detuerium fractionation has been best studied is IRAS 16293-2422, which was the subject of a recent multi-line HDO/H$_2$O study using Herschel/HIFI \citep{coutens12}.  This study found that water had considerably less deuterium fractionation than formaldehyde and methanol, and suggested that water formed in the pre-collapse phase while the organics were created later after CO depleted onto grains \citep{coutens12}.  In comets, the only deuterated organic species detected to date is DCN, which was seen in Hale-Bopp \citep{meier98b}.  These authors found the [DCN]/[HCN] ratio to be about an order of magnitude higher than the [HDO]/[H$_2$O] ratio \citep{meier98a}.  While DCN in Orion KL was not studied as a part of the present work, preliminary non-LTE modeling using the HIFI survey toward Orion KL as well as data from the 30 m IRAM telescope suggests [DCN]/[HCN] ratios in the Orion Hot Core and Compact Ridge between $(4-8) \times 10^{-3}$  (Marcelino et al., in preparation), in agreement with the ratios presented here and suggesting that HCN and water are similarly fractionated in Orion KL.

Table 8 shows that water, formaldehyde, and methanol all have lower deuterium fractionation in the Hot Core than in the Compact Ridge, although in some cases this is not significant within the quoted errors.  These two components are well known for having different chemical compositions, with oxygen-containing species more abundant toward the Compact Ridge and nitrogen-containing species more abundant toward the Hot Core.  The lower deuteration in the Hot Core is suggestive that the physical conditions in these two regions were different, with the Hot Core warmer than the Compact Ridge, in the evolutionary phase where the deuterium-rich organic material studied here was formed.

For all of these molecules, the combination of high molecular abundances and significant deuteration in warm gas is most consistent with synthesis on grain mantles \citep{rodgers96}.  The high [H$_2$D$^+$]/[H$_3$$^+$] ratio that arises at low temperatures from the slightly exothermic reaction of H$_3$$^+$ with HD also begets an enhanced atomic D/H ratio in the gas phase through dissociative recombination of H$_2$D$^+$ and other small deuterated molecules like DCO$^+$, and these D atoms can accrete onto grain surfaces and add to organic molecules.  Low-temperature gas phase ion-molecule chemistry has been generally found not to be able to produce the molecules in Table 8 at the observed abundances.  The observed deuterium fractionation, additionally, likely precludes gas-phase formation in the post-mantle evaporation phase ($T > 100$ K). Gas phase chemistry in the hot core phase could, however, reduce the deuterium fractionation, through, using ammonia as an example,

\begin{equation}
\textnormal{NH}_2\textnormal{D} + \textnormal{H}_3^+ \rightarrow \textnormal{NH}_3\textnormal{D}^+ + \textnormal{H}_2
\end{equation}

\noindent Other H-containing ions such as HCO$^+$ could also take the place of H$_3$$^+$ in this reaction.  The NH$_3$D$^+$ ion can cycle back to form ammonia through a electron recombination reaction, likely with an [NH$_2$D]/[NH$_3$] ratio near 3:1.  Over time, this chemisry will lower the D/H ratio, because there is no significant deuterium fractionation of H$_3$$^+$ or other deuterated ions at high temperatures to convert NH$_3$ into NH$_2$D.  It has been suggested that the reduction in the deuterium fractionation of organic species through this chemistry can occur in $\sim 10^3$--$10^4$ yr \citep{rodgers96}.

The deuterium fractionation of HDO and CH$_3$OD can also be affected by this chemistry, but not that of HDCO and CH$_2$DOH.  The lowest-energy form of protonated formaldehyde is H$_2$COH$^+$, a species that has been detected in Orion KL \citep{ohishi96}.  If HDCO is protonated, therefore, it yields the species HDCOH$^+$.  When this species is deprotonated, in order to produce formaldehyde, the same H atom from the protonation step must be removed, yielding HDCO again.  If the D atom is removed instead, this yields the carbene HCOH (hydroxymethylene).  This species has been observed to tunnel back to formaldehyde through a 15,000 K barrier in a few hours in cold ices \citep{schreiner08} but this process is likely much slower in the gas phase.  CH$_2$DOH, after protonation to form CH$_2$DOH$_2$$^+$, likewise can only yield CH$_2$DOH again after deprotonation.  On the other hand, when CH$_3$OD is protonated, yielding CH$_3$OHD$^+$, either CH$_3$OH or CH$_3$OD can be the end product.  \cite{osamura04} showed that the barriers for exchange of H and D between the methyl and hydroxyl positions in methanol are very high, whether in the neutral, ionic (CH$_3$OH$^+$ and isotopologues), or protonated ion forms of methanol.

In the Hot Core, the ALMA maps of HDO transitions presented in \cite{neill13} and of an NH$_2$D transition in Figure 3 confirm that the compact clumps in the Hot Core contain substantial abundances of deuterated molecules, and the detection of highly excited ($E_u > 500$ K) transitions of both species by HIFI confirms that this emission arises from very warm gas.  We observe about a factor of 2 higher deuteration in ammonia than in water, consistent within the statistical uncertainties with the factor of 1.5 predicted by statistics assuming equal deuteration between the two molecules \citep{rodgers96}.  \cite{jacq90} found similar results for a sample of hot cores, with NH$_2$D/NH$_3$ ratios consistently larger than HDO/H$_2$O, but the challenges in deriving reliable H$_2$O abundances (due to the lack of suitable transitions that can be observed from the ground) made this result unclear.

The four deuterated molecules observed in the Compact Ridge (HDO, HDCO, CH$_2$DOH, and CH$_3$OD) are found with similar abundances relative to their normal species.  Formaldehyde and methanol are believed to have coupled formation chemistry, resulting from sequential H additions to carbon monoxide on grain surfaces \citep{charnley97, rodgers02, watanabe02, hiraoka02, nagaoka05, fuchs09}:

\begin{equation}
\textnormal{CO} \xrightarrow{+\textnormal{H}, k_1} \textnormal{HCO} \xrightarrow{+ \textnormal{H}, k_2} \textnormal{H}_2\textnormal{CO} \xrightarrow{+\textnormal{H}, k_3} \textnormal{CH}_3\textnormal{O} \xrightarrow{+\textnormal{H}, k_4} \textnormal{CH}_3\textnormal{OH}
\end{equation}

\noindent The third step in this sequence could in principle go to CH$_2$OH instead, and that reaction is more exothermic, but also has been found to have a higher barrier and therefore CH$_3$O is believed to be the preferred product \citep{woon02}.  Any of these four steps can instead occur with a D atom, leading to deuterated analogs of formaldehyde and methanol: HDCO forms instead of H$_2$CO if either the first or second steps occurs with D instead of H; CH$_2$DOH forms if  deuterium adds at either the first, second, or third steps; and CH$_3$OD forms only through deuteration in the fourth step.  The first and third steps, being additions to closed-shell molecules, have activation barriers, while the second and fourth steps are believed to be barrierless.  The D/H branching ratios in the barrierless steps are likely to represent the atomic D/H accretion ratio \citep{rodgers02}, though this may not be true if back reactions (H abstraction) or competitive inhibition (i.e., if D or H react more quickly with another species on the grain) are significant.  The first and third steps, because they require tunneling through an energy barrier, will be highly dependent on the relative barriers and the grain temperature in addition to the relative D/H accretion rates.  Within this simple framework, a few limits can be established.  If each of the four steps is equally likely to proceed with H or D addition, we obtain statistical branching ratios: CH$_2$DOH/CH$_3$OD = 3, and HDCO has twice the fractionation of CH$_3$OD.  If, on the other hand, steps 1 and 3 add only hydrogen atoms and very little deuterium, and steps 2 and 4 have the same D/H branching ratio, HDCO, CH$_2$DOH and CH$_3$OD would all be found with the same degree of fractionation, interestingly in close agreement with the observations.  \cite{woon02} calculated that both the D + CO and D + H$_2$CO reactions have lower activation energy in a small ($\le 2$ molecule) water substrate; however, hydrogen atoms are more able to tunnel through barriers than deuterium, and tunneling is probably important in these processes at low temperatures.  Experiment and theory have found a strong kinetic isotope effect for the addition of H or D to CO, with the reaction with H proceeding faster at $T < 20$ K \citep{hidaka07, andersson11}.

The CH$_2$DOH/CH$_3$OD ratio of 1.2 derived here is consistent with the value previously derived by \cite{jacq93} for the Compact Ridge, but low-mass protostars have generally been found to have ratios much higher than the statistical value of 3 \citep{ratajczak11}.  Additionally, isotopologues of methanol with multiple deuterium atoms on the methyl group (CHD$_2$OH and CD$_3$OH) have been detected toward low-mass sources \citep{parise02, parise04, parise06}.  Therefore, theoretical and experimental work has been done to explore grain processes that could influence the deuterium branching ratios of formaldehyde and methanol.  \cite{nagaoka05} exposed pure solid CO ices to a beam of cold hydrogen and deuterium atoms with a D/H ratio of 0.1 (higher than the likely value for Orion KL, but consistent with the fractionation ratios seen toward low-mass protostars).  They detected deuterated formaldehyde and methanol, including multiply deuterated isotopologues, but no CH$_3$OD.  They also performed an experiment where they exposed pure solid CH$_3$OH to a beam of H and D atoms, and observed rapid D/H abstraction producing deuterated methanol, but again deuterium atoms were only incorporated into the methyl group.  This is likely due to the fact that the hydroxyl protons are more tightly bound in the ice matrix through hydrogen bonding than the methyl protons, so the hydroxyl protons are harder to abstract.  The dependence of this process on temperature and atomic D/H ratio (since it appears to be on the order of 0.01 in Orion KL) is not yet known.

There is also experimental evidence that methanol and water can exchange H and D atoms in ices \citep{ratajczak09}.  This process occurred at warmer temperatures ($\ge 120$ K) in a non-crystalline state.  Exchange in this process occurred only with the hydroxyl proton of methanol, rather than the methyl protons.  The equal deuterium fractionation of HDO and CH$_3$OD observed toward Orion KL makes this an intriguing scenario, but the physical conditions that could lead to this process are unknown.

Finally, as noted above, gas phase protonation/deprotonation reactions similar to equation (3) can alter the deuterium fractionation.  In general, as noted above, one would expect the abundances of HDO, NH$_2$D, and CH$_3$OD to be decreased by protonation/deprotonation chemistry relative to the hydrogenated species.  \cite{charnley97} proposed that with a very high HDO/H$_2$O ratio ($\sim 0.1$), in a few times $10^4$ years the CH$_2$DOH/CH$_3$OD ratio could be lowered through from the assumed initial value of 3 to nearly 1, due to the conversion of CH$_3$OH to CH$_3$OD through reactions with H$_2$DO$^+$.  However, as reported by \cite{neill13} and shown in Table 8, the observed HDO/H$_2$O ratio in the Compact Ridge is only about 0.005.  \cite{charnley97} also presented a model with HDO/H$_2$O = 0.01, close to the new observed value, and under these conditions the CH$_2$DOH/CH$_3$OD ratio only increased from 3 as CH$_3$OD was, on net, converted into CH$_3$OH while the relative abundance of CH$_2$DOH remained constant.  \cite{charnley97} also proposed that HDCO could be produced in the gas phase by the reaction of CH$_2$D with atomic oxygen.

The chemistry that influences the deuteration of these molecules is, as this discussion makes clear, complex, and our understanding many of the processes involved is limited.  Continued theoretical and experimental work will be needed to better understand the various possibilities.  The D/H ratios between the various chemical species suggest important differences between the chemistry of high-mass and low-mass star-forming regions.  Additionally, Orion KL may have unique chemistry; it has been proposed that the Hot Core is due to a recent ($\sim$ 500 years ago) explosive event \citep{zapata09, nissen12} that could have recently released the organic material of grain mantles into the gas phase \citep{zapata11}; the short amount of time that has elapsed since the putative formation event for these warm compact regions that evaporated the grain mantles might make the chemical content fresher than in regions where the hot phase chemistry is more evolved.  \cite{ratajczak11} did find similar upper limits on the CH$_2$DOH/CH$_3$OD ratios for two other high-mass star forming regions, W3(H$_2$O) ($< 0.9$) and G24.78+0.08 ($< 1.4$), although the CH$_3$OD line identification in G24.78+0.08 was uncertain due to an unusual velocity shift.  Further observations of deuterated species in other massive cores will help to reveal the extent to which the trends seen in Orion KL are representative of these sources as a whole.

\section{Conclusion}

We have presented abundances of deuterated forms of ammonia, formaldehyde, and methanol in the Orion KL Hot Core and Compact Ridge.  Due to the large number of transitions detected for each species and the wide energy and frequency coverage with homogenuous calibration, we consider the abundances and D/H ratios presented here to be the best constrained to date for any star forming region.  In the Hot Core, we find that ammonia has higher deuterium fractionation ($6.8 \times 10^{-3}$) than water ($3.0 \times 10^{-3}$) or methanol (upper limits of $4.2 \times 10^{-3}$ for CH$_2$DOH, and $1.8 \times 10^{-3}$ for CH$_3$OD).  In the Compact Ridge, we find similar deuteration in HDO, NH$_2$D, CH$_2$DOH, and CH$_3$OD, with a CH$_2$DOH/CH$_3$OD ratio of $1.2 \pm 0.3$, in agreement with the previous derivation of \cite{jacq93}, and clearly lower than the statistical ratio of 3 and the value observed toward low-mass protostars.  This could be due to the energetics of the steps in the formation pathway, or alternatively due to a number of other processes that can alter the relative abundances of the different isotopologues.  These abundances and D/H ratios can be used for comparison to other sources or to compare to chemical models of massive star forming regions.  A fuller understanding of the D/H ratios reported here will require further theoretical, experimental, and observational study.

\begin{acknowledgements}
HIFI has been designed and built by a consortium of institutes and university departments from across Europe, Canada, and the United States under the leadership of SRON Netherlands Institute for Space Research, Groningen, The Netherlands and with major contributions from Germany, France, and the US.  Consortium members are: Canada: CSA, U.Waterloo; France: CESR, LAB, LERMA, IRAM; Germany: KOSMA, MPIfR, MPS; Ireland: NUI Maynooth; Italy: ASI, IFSI-INAF, Osservatorio Astrofisico di Arcetri-INAF; Netherlands: SRON, TUD; Poland: CAMK, CBK; Spain: Observatorio Astron\'{o}mico Nacional (IGN), Centro de Astrobiolog\'{i}a (CSIC-INTA); Sweden: Chalmers University of Technology--MC2, RSS \& GARD, Onsala Space Observatory, Swedish National Space Board, Stockholm Observatory; Switzerland: ETH Zurich, FHNW; USA: Caltech, JPL, NHSC.  Support for this work was provided by NASA through an award issued by JPL/Caltech.  Author L.H.X thanks the Natural Sciences and Engineering Research Council of Canada for financial support of this research program.

This paper makes use of the following ALMA data:  ADS/JAO.ALMA\#2011.0.00009.SV.  ALMA is a partnership of ESO (representing its member states), NSF (USA) and NINS (Japan), together with NRC (Canada) and NSC and ASIAA (Taiwan), in cooperation with the Republic of Chile.  The Joint ALMA Observatory is operated by ESO, AUI/NRAO and NAOJ.

\end{acknowledgements}

\bibliography{OrionD}

\clearpage
\appendix

\begin{deluxetable}{c c c c l l l l}
\tablenum{2}
\tablewidth{0pt}
\tablecaption{Fit Gaussian parameters for detected transitions of NH$_2$D.}
\tablehead{ &Frequency & $E_u$ & $S_\textnormal{ij}\mu^2$ & $\int T_{mb}dv\tablenotemark{a}$ & $v_\textnormal{LSR}$ & $\Delta v_\textnormal{FWHM}$ & $N_u$\tablenotemark{b} \\
                   Transition & (MHz) & (K) & (D$^2$) &(K km s$^{-1}$) & (km s$^{-1}$) & (km s$^{-1}$) & ($10^{13}$ cm$^{-2}$)  }
\startdata
\emph{ortho} \\
$2_{02}$$^--1_{01}$$^-$ & 649916.4 & 47.7 & 0.94 & 2.12(24) & 6.56(15) & 6.4(5) & 80(9) \\
$2_{02}$$^--1_{10}$$^+$ & 512426.6 & 47.7 & 4.51 & 4.1(4) & 6.50(7) & 7.29(16) & 64(7) \\
$2_{11}$$^+-1_{10}$$^+$ & 717025.3 & 57.6 & 0.68 & 0.80(8) & 7.1(13) & 4.8(3) & 30(17)\\
$2_{21}$$^--1_{11}$$^+$ & 1074312.8 & 72.2 & 28.63 & 14.9(15) & 6.2(1) & 6.1(2) & 4.3(4) \\
$2_{20}$$^--1_{10}$$^+$ & 1038279.6 & 73.0 & 33.68 & 18.5(19) & 5.8(1) & 7.6(2) & 5.0(5) \\
$3_{03}$$^--2_{11}$$^+$ & 738699.1 & 93.0 & 5.03 & 3.1(3) & 6.17(8) & 5.43(19) & 21(2) \\
$3_{22}$$^--2_{12}$$^+$ & 1458010.3 & 120.1 & 31.80 & 7.8(18) & 4.2(7) & 5.3(12) & \emph{1.2(3)} \\
$3_{31}$$^+-3_{21}$$^-$ & 544942.1 & 149.6 & 13.27 & 2.1(5) & 7.1(5) & 5.5(14) & 13(3) \\
$3_{31}$$^+-2_{21}$$^-$ & 1612700.3 & 149.6 & 49.48 & 6.8(16) & 5.8(5) & 3.9(10) & \emph{0.52(12)} \\
$4_{04}$$^--3_{12}$$^+$ & 896707.9 & 151.6 & 4.14 & 3.8(11) & 6.2(7) & 4.9(17) & 23(7) \\
$4_{14}$$^+-3_{22}$$^-$ & 657607.0 & 151.7 & 2.32 & 1.04(22) & 6.4(4) & 4.8(11) & 28(6) \\
$4_{13}$$^+-3_{21}$$^-$ & 1081841.9 & 175.4 & 8.91 & 7.1(7) & 6.45(8) & 5.27(20) & 11.5(12) \\
$4_{23}$$^--3_{31}$$^+$ & 702480.5 & 183.3 & 1.71 & 0.77(8) & 6.4(2) & 2.8(4) & 23(2) \\
$4_{23}$$^--3_{13}$$^+$ & 1863397.6 & 183.3 & 34.40 & 15.4(25) & 5.6(4) & 6.3(15)& \emph{1.5(2)} \\
$4_{22}$$^--3_{30}$$^+$ & 877406.4 & 191.9 & 2.31 & 1.35(21) & 5.84(28) & 4.9(7) & 15.5(24) \\
$4_{22}$$^--4_{14}$$^+$ & 838041.4 & 191.9 & 4.08 & 0.85(12) & 6.31(17) & 3.32(30) & 6.3(9) \\
$4_{22}$$^--3_{12}$$^+$ & 1735396.5 & 191.9 & 58.09 & 11.3(18) & 6.6(4) & 5.1(8) & \emph{0.8(1)} \\
$4_{31}$$^--4_{23}$$^-$ & 671117.3 & 215.6 & 16.71 & 3.3(4) & 6.17(14) & 5.6(4) & 11.3(13) \\
$5_{05}$$^--4_{13}$$^+$ & 1002366.5 & 223.5 & 3.42 & 1.12(19) & 7.27(24) & 3.3(5) & 7.2(12) \\
$4_{41}$$^--4_{31}$$^+$ & 834183.6 & 255.6 & 13.37 & 1.7(9) & 6.8(8) & 3.2(20) & 3.9(20) \\
$4_{40}$$^--4_{32}$$^+$ & 855075.8 & 255.6 & 13.20 & 3.6(5) & 5.6(4) & 6.7(10) & 7.8(12) \\
$5_{24}$$^--4_{32}$$^+$ & 976894.5 & 261.5 & 3.62 & 1.13(18) & 7.34(26) & 4.0(6) & 7.4(12) \\
$5_{32}$$^+-5_{24}$$^-$ & 785598.0 & 299.2 & 15.22 & 3.1(4) & 6.4(3) & 7.6(10) & 9.3(13) \\
$6_{16}$$^+-5_{24}$$^-$ & 973410.3 & 308.2 & 2.90 & 1.11(21) & 7.9(3) & 4.1(7) & 10.9(21) \\
$5_{42}$$^--5_{32}$$^+$ & 785703.3 & 336.9 & 23.53 & 1.27(20) & 6.65(21) & 3.4(5) & 2.4(4) \\
$5_{41}$$^--5_{33}$$^+$ & 862285.3 & 337.1 & 22.14 & 1.71(26) & 6.7(3) & 4.4(6) & 2.6(4) \\
$6_{25}$$^--5_{33}$$^+$ & 1214306.3 & 354.0 & 5.07 & 0.75(2) & 7.1(8) & 4.3(11) & 2.3(8) \\
$5_{50}$$^+-5_{42}$$^-$ & 1067346.6 & 388.1 & 5.07 & 3.0(4) & 6.54(22) & 6.0(7) & 4.0(5) \\
$6_{33}$$^+-6_{25}$$^-$ & 983372.5 & 401.2 & 13.70 & 2.6(4) & 6.0(3) & 6\tablenotemark{c} & 6.6(11) \\
$6_{43}$$^--6_{33}$$^+$ & 698717.6 & 434.7 & 10.62 & 1.58(27) & 5.2(4) & 5.5(8) & 3.6(6)\\
$7_{53}$$^+-7_{43}$$^-$ & 985711.9 & 599.4 & 33.14 & 1.49(15) & 7.27(15) & 5.86(15) & 1.4(1) \\
$7_{52}$$^+-7_{44}$$^-$ & 1055703.6 & 599.7 & 32.06 & 0.80(9) & 7.37(14) & 2.5(3) & 0.65(7) \\
\hline
\emph{para} \\
$1_{10}$$^--0_{00}$$^+$ & 494454.5 & 23.7 & 6.37 & 3.6(18) & 6.5(22) & 6.7(50) & 9(5) \\
$2_{02}$$^+-1_{01}$$^+$ & 649955.8 & 47.2 & 0.30 & 0.55(17) & 6(7) & 4.7(15) & 22(7) \\
$2_{02}$$^+-1_{10}$$^-$ & 488323.8 & 47.2 & 1.50 & 0.48(14) & 7.5(9) & 5\tablenotemark{c} & 8.8(26) \\
$2_{21}$$^+-1_{11}$$^-$ & 1050452.4 & 71.7 & 9.54 & 12.1(5) & 6.31(14) & 7.1(4) & 3.7(4) \\
$3_{12}$$^--2_{20}$$^+$ & 765633.2 & 109.2 & 1.34 & 0.8(5) & 6.5(13) & 4.3(27) & 6.1(35)\\
$3_{21}$$^+-3_{13}$$^-$ & 592363.8 & 122.9 & 2.21 & 0.8(5) & 6.5(22) & 6.5(53) & 8(5)\\
$3_{31}$$^--2_{21}$$^+$ & 1635742.5 & 150.2 & 16.49 & 13.3(19) & 5.6(3) & 5\tablenotemark{c} & \emph{0.98(14)} \\
$3_{30}$$^--3_{22}$$^+$ & 641057.6 & 150.3 & 3.95 & 1.68(20) & 6.31(20) & 6.2(5) & 7.3(9) \\
$4_{04}$$^+-3_{12}$$^-$ & 872891.9 & 151.1 & 1.37 & 0.56(12) & 7.1(4) & 3.7(8) & 3.7(8) \\
$4_{13}$$^--3_{21}$$^+$ & 1104757.1 & 175.9 & 2.98 & 2.4(4) & 6.0(4) & 4.8(9) & 3.7(7) \\
$4_{13}$$^--3_{03}$$^+$ & 1740163.2 & 175.9 & 12.13 & 8.7(28) & 6.1(8) & 4.6(16) & \emph{0.95(30)} \\
$4_{22}$$^+-4_{14}$$^-$ & 814702.2 & 191.4 & 1.36 & 0.68(14) & 7.4(6) & 6.6(11) & 5.5(11) \\
$4_{32}$$^--4_{22}$$^+$ & 494928.7 & 215.1 & 8.20 & 1.32(14) & 6.23(18) & 7.7(4) & 7.6(8) \\
$4_{31}$$^--4_{23}$$^+$ & 693656.4 & 216.1 & 5.58 & 1.36(17) & 6.27(24) & 6.1(7) & 4.3(5) \\
$5_{41}$$^+-5_{33}$$^-$ & 841031.5 & 336.6 & 7.37 & 1.60(14) & 7.1(3) & 7.7(7) & 2.7(2) \\
\enddata
\tablenotetext{a}{Uncertainties for line fluxes have been calculated assuming a 10\% calibration uncertainty.}
\tablenotetext{b}{Calculated using equation (1), assuming optically thin lines.  Lines in bands 6-7 (1425-1906 GHz) are not used due to optical depth (see text).}
\tablenotetext{c}{This parameter was held constant in the fit.}
\end{deluxetable}

\begin{deluxetable}{c c c l l l l l l l l l}
\tablenum{3}
\rotate
\tablewidth{0pt}
\tabletypesize{\scriptsize}
\tablecaption{Fit Gaussian parameters for detected transitions of H$_2$$^{12}$CO.}
\tablehead{ & & & \multicolumn{3}{c}{Compact Ridge} & \multicolumn{3}{c}{Plateau} & \multicolumn{3}{c}{Hot Core} \\
                   & Frequency & $E_u$ & $\int T_{mb}dv\tablenotemark{a}$ & $v_\textnormal{LSR}$ & $\Delta v_\textnormal{FWHM}$ & $\int T_{mb}dv\tablenotemark{a}$ & $v_\textnormal{LSR}$ & $\Delta v_\textnormal{FWHM}$ & $\int T_{mb}dv\tablenotemark{a}$ & $v_\textnormal{LSR}$ & $\Delta v_\textnormal{FWHM}$  \\
                   Transition & (MHz) & (K) & (K km s$^{-1}$) & (km s$^{-1}$) & (km s$^{-1}$) & (K km s$^{-1}$) & (km s$^{-1}$) & (km s$^{-1}$) & (K km s$^{-1}$) & (km s$^{-1}$) & (km s$^{-1}$) }
\startdata
$2_{02}-1_{01}$ & 145602.9 & 10.4 & 65(7) & 8.0(1) & 4.55(3) & 31(4) & 4.6(4) & 31.3(17) & 26.5(29) & 6.9(2) & 12.8(8) \\
$2_{12}-1_{11}$ & 140839.5 & 21.9 & 70(8) & 7.6(1) & 5.31(24) & 25(3) & 13.6(11) & 35(13) & 36(12) & 4.7(22) & 18(6) \\
$2_{11}-1_{10}$ & 150498.3 & 22.6 & 54(6) & 8.0(1) & 4.23(9) & 53(7) & 6.8(2) & 25.7(11) & 33(5) & 7.7(1) & 9.4(8) \\
$7_{16}-7_{17}$ & 135030.4 & 112.8 & 4.9(7) & 8.0(1) & 4.6(4) & & & & 3.2(7) & 5.1(16) & 18(4) \\
$7_{07}-6_{06}$ & 505833.7 & 97.4 & 33(3) & 9.06(1) & 4.10(2) & 52(5) & 7.59(1) & 25\tablenotemark{c} & 42(4) & 7.75(1) & 11.6(1) \\
$7_{16}-6_{15}$ & 525665.8 & 112.8  & 51(5) & 9.24(1) & 4.73(1) & 117(12) & 9.12(2) & 23.8(5) & 28(3) & 6.22(2) & 11.15(5) \\
$8_{08}-7_{07}$ & 576708.3 & 125.1 & 36(4) & 8.94(1) & 4.51(1) & 65(7) & 7.60(4) & 26.1(1) & 46(5) & 5.49(1) & 13.73(8) \\
$8_{18}-7_{17}$ & 561899.3 & 133.3 & 48(5) & 9.25(1) & 4.46(2) & 119(12) & 9.19(4) & 25.68(8) & 63(6) & 7.48(3) & 12.49(10) \\
$8_{17}-7_{16}$ & 600330.6 & 141.6 & 52(5) & 9.1(3) & 4.9(3) & 132(13) & 8.3(3) & 24.3(3) & 24.0(26) & 5.1(3) & 9.2(3) \\
$7_{25}-6_{24}$ & 513076.4 & 145.4 & 32(3) & 8.86(1) & 4.61(3) & 45(5) & 8.67(9) & 25\tablenotemark{c} & 19.0(19) & 5.2\tablenotemark{c} & 10\tablenotemark{c} \\
$9_{09}-8_{08}$ & 647081.7 & 156.2 & 47(5) & 8.89(23) & 5.0(23) & 83(8) & 7.9(23) & 24.07(23) & 33(3) & 5.1(23) & 10.3(3) \\
$8_{27}-7_{26}$ & 581611.8 & 172.8 & 23.7(24) & 8.70(1) & 4.13(2) & 42(5) & 7.25(11) & 19.7(3) & 28(3) & 6.47(2) & 11.1(3) \\
$10_{19}-9_{18}$ & 749071.9 & 209.9 & 26.1(26) & 8.9(2) & 3.9(2) & 108(11) & 8.2(2) & 24.4(2) & 67(7) & 6.6(2) & 11.2(2) \\
$7_{35}-6_{34}$ & 510155.7 & 203.9 & 29.9(29) & 9.01(9) & 4.3(3) & 64(6) & 7.7(6) & 25\tablenotemark{c} & 30.3(30) & 7.1(6) & 10.4(10) \\
$7_{34}-6_{33}$ & 510237.8 & 203.9 & 30.2(30) & 9.01(1) & 4.34(2) & 59(6) & 8.22(6) & 25\tablenotemark{c} & 37.1(37) & 6.89(4) & 11.67(9) \\
$11_{0\, 11}-10_{0\, 10}$ & 786285.0 & 228.3 & 26.2(26) & 8.82(1) & 4.88(4) & 76(8) & 6.25(8) & 26.1(4) & 46(5) & 6.84(1) & 10.82(20)\\
$8_{36}-7_{35}$ & 583144.6 & 231.9 & 33(4) & 8.91(3) & 4.66(9) & 73(7) & 9.45(1) & 24.1(4) & 28(4) & 5.7(3) & 10.8(3) \\
$8_{35}-7_{34}$ & 583308.6 & 231.9 & 26.4(26) & 9.02(26) & 4.28(26) & 66(7) & 9.17(25) & 25\tablenotemark{c} & 40.5(9) & 6.51(26) & 10.13(26) \\
$10_{29}-9_{28}$ & 726208.3 & 239.1 & 16.3(17) & 8.4(2) & 3.8(2) & 37(4) & 6.6(2) & 23.6(2) & 44(4) & 6.3(2) & 10.7(2) \\
$10_{28}-9_{27}$ & 737342.7 & 240.7 & 14.6(15) & 8.3(2) & 3.8(2) & 40(4) & 9\tablenotemark{c} & 25\tablenotemark{c} & 36(4) & 5.9(2) & 10.0(2) \\
$11_{1\, 10}-10_{19}$ & 823082.8 & 249.4 & 26.5(27) & 8.69(2) & 4.56(4) & 76(8) & 8.39(2) & 24.18(14) & 70(7) & 6.28(2) & 12.28(8) \\
$12_{0\, 12}-11_{0\, 11}$ & 855151.3 & 269.4 & 19(4) & 8.05(7) & 4.5(5) & 23(4) & 8.0(11) & 22.8(23) & 46(8) & 5.4(5) & 12.2(5) \\
$9_{36}-8_{35}$ & 656464.6 & 263.4 & 32(3) & 8.60(1) & 4.70(4) & 57(6) & 9.17(1) & 22.0(2) & 43(5) & 5.48(11) & 11.97(11) \\
$11_{29}-10_{28}$ & 812831.4 & 279.7 & 13.2(14) & 8.04(1) & 3.90(5) & 26.6(28) & 8.7(2) & 26.0(5) & 41(4) & 5.95(5) & 11.5(13) \\
$7_4-6_4$\tablenotemark{b} & 509829.6 & 286.2 & 15.6(23) & 8.28(5) & 4.0(3) & 11\tablenotemark{c} & 9\tablenotemark{c} & 25\tablenotemark{c} & 26(3) & 6.2(4) & 12.3(7) \\
$10_{38}-9_{37}$ & 729212.5 & 298.4 & 25.8(26) & 8.48(1) & 4.61(4) & 56\tablenotemark{c} & 9\tablenotemark{c} & 25\tablenotemark{c} & 47(5) & 5.66(4) & 11.43(6) \\
$10_{37}-9_{36}$ & 729725.0 & 298.4 & 25.3 (31) & 8.35(6) & 4.56(20) & 15\tablenotemark{c} & 9\tablenotemark{c} & 25\tablenotemark{c} & 75(8) & 6.04(13) & 14.2(3) \\
$13_{1\, 13}-12_{1\, 12}$ & 909507.7 & 318.2 & 21.8(33) & 8.36(11) & 4.7(3) & 46(5) & 8.73(3) & 21.4(11) & 64(7) & 5.5(2) & 12.0(4) \\
$12_{2\, 11}-11_{2\, 10}$ & 870273.5 & 319.2 & 11.1(21) & 7.87(11) & 4.0(4) & 17(8) & 7.6(4) & 22(4) & 42(9) & 5.6(3) & 11.9(7) \\
$8_4-7_4$ & 582723.0 & 314.1 & 18.9(19) & 7.91(1) & 4.67(5) & 11\tablenotemark{c} & 9\tablenotemark{c} & 25\tablenotemark{c} & 31(3) & 5.16(7) & 13.49(12) \\
$9_4-8_4$ & 655639.9 & 345.6 & 18.9(23) & 7.05(6) & 5.1(2) & 12\tablenotemark{c} & 9\tablenotemark{c} & 25\tablenotemark{c} & 41(4) & 4.6(2) & 13.8(4) \\
$12_{3\, 10}-11_{39}$ & 875366.2 & 378.9 & 13.9(15) & 7.91(3) & 4.04(9) & 11\tablenotemark{c} & 9\tablenotemark{c} & 25\tablenotemark{c} & 65(7) & 5.92(3) & 12.87(9) \\
$12_{39}-11_{38}$ & 876649.1 & 379.0 & 16(5) & 8.0(2) & 4.5(8) & 18\tablenotemark{c} & 9\tablenotemark{c} & 25\tablenotemark{c} & 58(7) & 5.6(3) & 12.8(7) \\
$14_{1\, 13}-13_{1\, 12}$ & 1043222.9 & 389.1 & 16.9(22) & 7.61(6) & 5.1(2) & 24\tablenotemark{c} & 9\tablenotemark{c} & 25\tablenotemark{c} & 63(6) & 5.12(8) & 13.2(2) \\
$15_{0\, 15}-14_{0\, 14}$ & 1059441.7 & 412.1 & 19(3) & 6.8(1) & 5.6(3) & 18\tablenotemark{c} & 9\tablenotemark{c} & 25\tablenotemark{c} & 27(4) & 1.8(6) & 12(6) \\
$13_{3\, 10}-12_{39}$ & 950364.9 & 424.6 & 11.9(15) & 7.65(6) & 4.3(2) & 11\tablenotemark{c} & 9\tablenotemark{c} & 25\tablenotemark{c} & 53(5) & 5.40(9) & 12.0(1)\\
$15_{1\, 14}-14_{1\, 13}$ & 1115832.9 & 442.7 & 14(5) & 7.1(3) & 5.2(8) & & & & 59(7) & 5.1(2) & 13.5(8) \\
$8_5-7_5$ & 582382.1 & 419.8 & 13.8(16) & 8.25(4) & 4.36(14) & 17\tablenotemark{c} & 9\tablenotemark{c} & 25\tablenotemark{c} & 33(3) & 6.31(11) & 13.2(2) \\
$15_{2\, 14}-14_{2\, 13}$ & 1085167.6 & 465.1 & 4.8(13) & 7.02(16) & 3.8(5) & & & & 28(3) & 5.24(14) & 10.4(3) \\
$16_{1\, 16}-15_{1\, 15}$ & 1116331.2 & 469.1 & 10.6(16) & 7.51(8) & 4.8(3) & 14\tablenotemark{c} & 9\tablenotemark{c} & 25\tablenotemark{c} & 55(6) & 4.85(9) & 11.86(14) \\
$9_5-8_5$ & 655212.1 & 451.2 & 14.7(18) & 7.96(5) & 4.4(2) & 14\tablenotemark{c} & 9\tablenotemark{c} & 25\tablenotemark{c} & 40(4) & 5.52(14) & 12.9(3) \\
$14_{3\, 12}-13_{3\, 11}$ & 1021533.1 & 473.4 & 10.1(15) & 7.38(8) & 4.3(3) & 14\tablenotemark{c} & 9\tablenotemark{c} & 25\tablenotemark{c} & 42(4) & 4.72(11) & 11.5(2)\\
$14_{3\, 11}-13_{3\, 10}$ & 1024288.6 & 473.8 & 19(5) & 7.2(2) & 5.8(7) & 9\tablenotemark{c} & 9\tablenotemark{c} & 25\tablenotemark{c} & 41(6) & 4.2(5) & 12.9(6) \\
$16_{1\, 15}-15_{1\, 14}$ & 1187985.7 & 499.7 & 16(7) & 6.9(3) & 6.6(9) & & & & 29(7) & 4.9(5) & 12.5(12) \\
$10_5-9_5$ & 728053.5 & 486.2 & 13.8(18) & 7.89(6) & 4.6(2) & 20\tablenotemark{c} & 9\tablenotemark{c} & 25\tablenotemark{c} & 36(4) & 5.2(1) & 12.0(2) \\
$16_{2\, 15}-15_{2\, 14}$ & 1156435.5 & 520.6 & 8.2(9) & 6.8(2) & 5.4(3) & & & & 14.5(16) & 3.56(8) & 10.7(5) \\
$17_{1\, 17}-16_{1\, 16}$ & 1184994.4 & 525.9 & & & & & & & 29.1(30) & 4.81(8) & 8.37(15) \\
$15_{3\, 13}-14_{3\, 12}$ & 1094589.8 & 526.0 & 7.0(8) & 7.21(5) & 3.86(17) & & & & 43(4) & 5.10(5) & 11.24(10) \\
$16_{2\, 14}-15_{2\, 13}$ & 1193363.5 & 529.5 & & & & & & & 21.8(22) & 5.37(1) & 9.58(12) \\
$8_6-7_6$ & 582070.8 & 548.7 & 2.6(4) & 7.35(7) & 4.6(2) & & & & 6.1(7) & 5.18(17) & 11.0(3) \\
$17_{1\, 16}-16_{1\, 15}$ & 1259639.9 & 560.1 & 10.1(16) & 6.4(1) & 5.0(3) & & & & 38(4) & 4.38(12) & 13.5(4) \\
$12_5-11_5$ & 873775.4 & 566.5 & 6.7(7) & 7.55(17) & 3.92(17) & & & & 46(5) & 5.47(17) & 12.5(17) \\
$17_{2\, 16}-16_{2\, 15}$ & 1227506.6 & 579.5 & & & & & & & 11.8(13) & 5.98(18) & 8.7(5) \\
$16_{3\, 14}-15_{3\, 13}$ & 1167608.5 & 582.0 & 5.2(9) & 7.09(9) & 3.8(3) & & & & 29(3) & 5.19(10) & 9.99(18) \\
$16_{3\, 13}-15_{3\, 12}$ & 1172882.9 & 582.8 & 8.9(11) & 6.64(7) & 4.8(2) & & & & 26.8(28) & 4.10(13) & 13.2(3) \\
$18_{0\, 18}-17_{0\, 17}$ & 1261739.9 & 584.1 & & & & & & & 13.6(15) & 5.10(13) & 6.0(3) \\
$17_{2\, 15}-16_{2\, 14}$ & 1269483.0 & 590.4 & & & & & & & 15.5(17) & 5.45(19) & 7.4(5) \\
$9_6-8_6$ & 654838.2 & 580.2 & & & & & & & 13.4(13) & 6.88(4) & 9.42(9) \\
$15_{4\, 12}-14_{4\, 11}$ & 1093759.7 & 608.0 & & & & & & & 13.9(14) & 5.54(5) & 7.57(13) \\
$15_{4\, 11}-14_{4\, 10}$ & 1093914.9 & 608.0 & & & & & & & 15.9(16) & 5.20(5) & 8.43(13) \\
$13_5-12_5$ & 946658.4 & 612.0 & & & & & & & 26.5(27) & 6.08(12) & 8.9(3) \\
$10_6-9_6$ & 727608.7 & 615.1 & & & & & & & 15.8(16) & 5.83(3) & 9.40(8) \\
$17_{3\, 15}-16_{3\, 14}$ & 1240572.2 & 641.5 & & & & & & & 23.4(23) & 5.43(21) & 7.8(5) \\
$17_{4\, 13}-16_{4\, 12}$ & 1240431.1 & 723.5 & & & & & & & 5.9(6) & 5.6(15) & 6.7(4) \\
$8_7-7_7$ & 581750.0 & 700.9 & & & & & & & 4.0(4) & 6.21(17) & 6.8(4) \\
$16_5-15_5$ & 1165408.0 & 769.3 & & & & & & & 24.4(24) & 4.88(7) & 9.48(19) \\
$10_7-9_7$ & 727174.8 & 767.1 & & & & & & & 11.6(12) & 5.62(5) & 8.81(14) \\
$11_7-10_7$ & 799884.1 & 805.6 & & & & & & & 11.7(12) & 5.1(2) & 9.1(6) \\
$17_5-16_5$ & 1238362.3 & 828.7 & & & & & & & 19.5(20) & 3.83(8) & 8.73(18) \\
$16_6-15_6$ & 1164311.6 & 898.0 & & & & & & & 5.8(6) & 5.5(5) & 6.5(11) \\
$17_6-16_6$ & 1237111.7 & 957.3 & & & & & & & 3.3(3) & 5.9(6) & 5.5(12) \\
$15_7-14_7$ & 1090695.0 & 994.0 & & & & & & & 4.9(5) & 5.5(4) & 5.5(8) \\
$12_8-11_8$ & 872016.1 & 1022.5 & & & & & & & 1.32(13) & 4.9(7) & 5.6(15) \\
\enddata
\tablenotetext{a}{Uncertainties for line fluxes have been calculated assuming a 10\% calibration uncertainty.}
\tablenotetext{b}{Where only one $K$ quantum number is given, two transitions (e.g., $7_{44}-6_{43}$ and $7_{43}-6_{42}$) are degenerate to within the HIFI frequency resolution.}
\tablenotetext{c}{Denotes that this parameter was held constant in the fit, or was fit by eye.}
\end{deluxetable}

\begin{deluxetable}{c c c l l l l l l}
\tablenum{4}
\tablewidth{0pt}
\tabletypesize{\footnotesize}
\tablecaption{Fit Gaussian parameters for detected transitions of H$_2$$^{13}$CO.}
\tablehead{ & & & \multicolumn{3}{c}{Compact Ridge} & \multicolumn{3}{c}{Hot Core} \\
                   & Frequency & $E_u$ & $\int T_{mb}dv\tablenotemark{a}$ & $v_{LSR}$ & $\Delta v_{FWHM}$ & $\int T_{mb}dv\tablenotemark{a}$ & $v_{LSR}$ & $\Delta v_{FWHM}$  \\
                   Transition & (MHz) & (K) & (K km s$^{-1}$) & (km s$^{-1}$) & (km s$^{-1}$) & (K km s$^{-1}$) & (km s$^{-1}$) & (km s$^{-1}$)  }
\startdata
$2_{02}-1_{01}$ & 141983.7 & 10.2 & 3.1(3) & 7.92(4) & 3.64(8) & 1.69(19) & 5.5(5) & 12.4(11) \\
$2_{12}-1_{11}$ & 137450.0 & 21.7 & 7.8(8) & 7.79(5) & 4.67(26) & 1.4(8) & 6.1(35) & 14(5) \\
$2_{11}-1_{10}$ & 146635.7 & 22.4 & 6.4(6) & 7.76(6) & 3.86(21) & 1.60(24) & 3.0(8) & 15.3(20) \\
$7_{07}-6_{06}$ & 493531.1 & 95.0 & 2.8(3) & 8.33(4) & 3.67(14) & 1.12(22) & 4.9(7) & 9.0(8)\\
$7_{17}-6_{16}$ & 480194.7 & 104.1 & 5.0(5) & 8.44(3) & 3.29(8) & 8.7(9)) & 6.58(16) & 12.9(3) \\
$7_{16}-6_{15}$ & 512250.4 & 110.3 & 6.1(6) & 8.25(2) & 3.77(5) &  4.71(5) & 5.31(7) & 11.1(3) \\
$8_{08}-7_{07}$ & 562777.6 & 122.1 & 1.66(20) & 8.13(3) & 3.00(14) & 3.8(4) & 5.90(15) & 10.2(3) \\
$8_{18}-7_{17}$ & 548475.1 & 130.4 & 4.2(5) & 8.167(22) & 3.26(9) & 7.2(7) & 6.37(11) & 9.62(18) \\
$8_{17}-7_{16}$ & 585041.1 & 138.3 & 5.4(6) & 8.06(6) & 4.52(16) & 6.8(7) & 4.92(11) & 13.1(5) \\
$7_{26}-6_{25}$ & 496515.1 & 142.7 & 0.96(11) & 8.21(6) & 3\tablenotemark{b} & 1.0(14) & 7.03(25) & 7.4(5) \\
$9_{19}-8_{18}$ & 616638.9 & 160.0 & 2.6(3) & 8.02(2) & 2.75(9) & 9.2(9) & 6.71(6) & 10.7(2) \\
$9_{18}-8_{17}$ & 657665.2 & 169.9 & 2.8(6) & 7.73(10) & 3.2(4) & 8.0(9) & 5.7(3) & 11.8(7) \\
$10_{0\, 10}-9_{09}$ & 699874.8 & 186.0 & 0.8(4) & 7.7(3) & 2.3(7) & 1.3(5) & 4.7(1) & 5.0(13) \\
$10_{1\, 10}-9_{19}$ & 684677.9 & 192.9 & 2.4(3) & 7.63(6) & 2.98(23) & 10.1(11) & 6.07(16) & 12.1(4) \\
$11_{1\, 11}-10_{1\, 10}$ & 752585.9 & 229.0 & 1.48(16) & 7.2(5) & 3\tablenotemark{b} & 8.5(9) & 6.72(10) & 11.8(3) \\
$10_{29}-9_{28}$ & 708269.2 & 234.5 & & & & 1.07(11) & 6.34(16) & 5.4(3) \\
$10_{28}-9_{27}$ & 718352.2 & 236.0 & & & & 2.82(28) & 6.20(21) & 8.2(6) \\
$12_{1\, 11}-11_{1\, 10}$ & 874246.5 & 285.4 & & & & 3.12(31) & 6.34(9) & 5.65(21) \\
$13_{1\, 13}-12_{1\, 12}$ & 887994.3 & 311.0 & & & & 4.3(4) & 5.76(13) & 6.75(31) \\
$11_{39}-10_{38}$ & 782212.5 & 331.5 & & & & 2.31(23) & 6.43(11) & 5.79(28) \\
$11_{38}-10_{37}$ & 782921.2 & 331.6 & & & & 3.13(31) & 6.34(22) & 7.4(5) \\
$13_{2\, 11}-12_{2\, 10}$ & 939543.3 & 360.6 & & & & 1.00(10) & 5.26(25) & 5.3(6)  \\
$13_{3\, 11}-12_{3\, 10}$ & 924714.9 & 416.8 & & & & 2.30(23) & 6.38(16) & 5.54(16) \\
$13_{3\, 10}-12_{39}$ & 926350.4 & 417.1 & & & & 1.49(15) & 7.4(4) & 7.6(6) \\
$8_5-7_5$\tablenotemark{c} & 567863.2 & 417.8 & & & & 0.77(8) & 7.05(24) & 6.6(5) \\
$12_{49}-11_{48}$ & 852689.7 & 454.9 & & & & 1.43(14) & 8.16(10) & 4.08(19) \\
$10_5-9_5$ & 709892.2 & 482.5 & & & & 2.74(27) & 7.6(5) & 7\tablenotemark{b} \\
\enddata
\tablenotetext{a}{Uncertainties for line fluxes have been calculated assuming a 10\% calibration uncertainty.}
\tablenotetext{b}{Denotes that this parameter was held constant in the fit.}
\tablenotetext{c}{Where only one $K$ quantum number is given, two transitions (e.g., $7_{44}-6_{43}$ and $7_{43}-6_{42}$) are degenerate to within the HIFI frequency resolution.}
\end{deluxetable}

\begin{deluxetable}{c c c l l l l l l}
\tablenum{5}
\tablewidth{0pt}
\tabletypesize{\footnotesize}
\tablecaption{Fit Gaussian parameters for detected transitions of HDCO.}
\tablehead{ & & & \multicolumn{3}{c}{Compact Ridge} & \multicolumn{3}{c}{10.4 km s$^{-1}$ component} \\
                   & Frequency & $E_u$ & $\int T_{mb}dv\tablenotemark{a}$ & $v_{LSR}$ & $\Delta v_{FWHM}$ & $\int T_{mb}dv\tablenotemark{a}$ & $v_{LSR}$ & $\Delta v_{FWHM}$  \\
                   Transition & (MHz) & (K) & (K km s$^{-1}$) & (km s$^{-1}$) & (km s$^{-1}$) & (K km s$^{-1}$) & (km s$^{-1}$) & (km s$^{-1}$)  }
\startdata
$2_{11}-1_{01}$ & 134284.8 & 17.6 & 2.6(3) & 8.13(15) & 5.0(4) \\
$9_{0 9}-8_{0 8}$  & 565857.529 &  137.28 &  2.4(4)  &  8.06(23)  &  2.9(5)  &  0.5(3)  &  10.8(6)  &  2.0(9) \\
$9_{1 9}-8_{1 8}$  & 552740.898 &  140.97 &  1.87(8)  &  7.784(20)  &  3.3(6)  &  1.20(15)  &  9.9(4)  &  4.2(8) \\
$10_{0 10}-9_{0 9}$ & 625687.651 &  167.30 &  2.0(10)  &  7.8\tablenotemark{a}  &  3.8(22)  &  1.7(6)  &  10.4\tablenotemark{a}  &  4.5(21) \\
$9_{2 7}-8_{2 6}$ & 591898.719 & 172.70 &  1.79(21)  &  7.8\tablenotemark{a}  &  3.5(5)  &  0.58(14)  &  10.7(3)  &  2.5\tablenotemark{a} \\
$11_{0 11}-10_{0 10}$ & 684995.747 & 200.17 & 1.38(9) & 7.8\tablenotemark{a} & 3.0\tablenotemark{a} &  0.50(9) & 10.43(24)  & 2.5(5)    \\
$9_{3 7}-8_{3 6}$ & 581677.779 & 211.05 &  0.5(4) & 8.3(15)  &  3.7(33) \\
$10_{3 8}-9_{3 7}$ & 646525.308 & 242.08 & 0.82(9) & 7.2(19) & 3.9(5) & 0.23(7) & 10.96(25) & 1.9(5) \\ 
$10_{3 7}-9_{3 6}$ & 648030.611 & 242.24 & 0.90(20) & 7.8\tablenotemark{a} & 3.0\tablenotemark{a} & 0.22(18) & 10.4\tablenotemark{a} & 2.5\tablenotemark{a} \\
$11_{3 8}-10_{3 7}$ & 713801.243 & 276.50 & 0.55(6) & 7.8\tablenotemark{a} & 3.0\tablenotemark{a} & 0.22(6) & 10.4\tablenotemark{a} & 2.5\tablenotemark{a}
\enddata
\tablenotetext{a}{Uncertainties for line fluxes have been calculated assuming a 10\% calibration uncertainty.}
\tablenotetext{b}{Denotes that this parameter was held constant in the fit.}
\end{deluxetable}

\begin{figure}
\figurenum{10}
\includegraphics[width=6.0in]{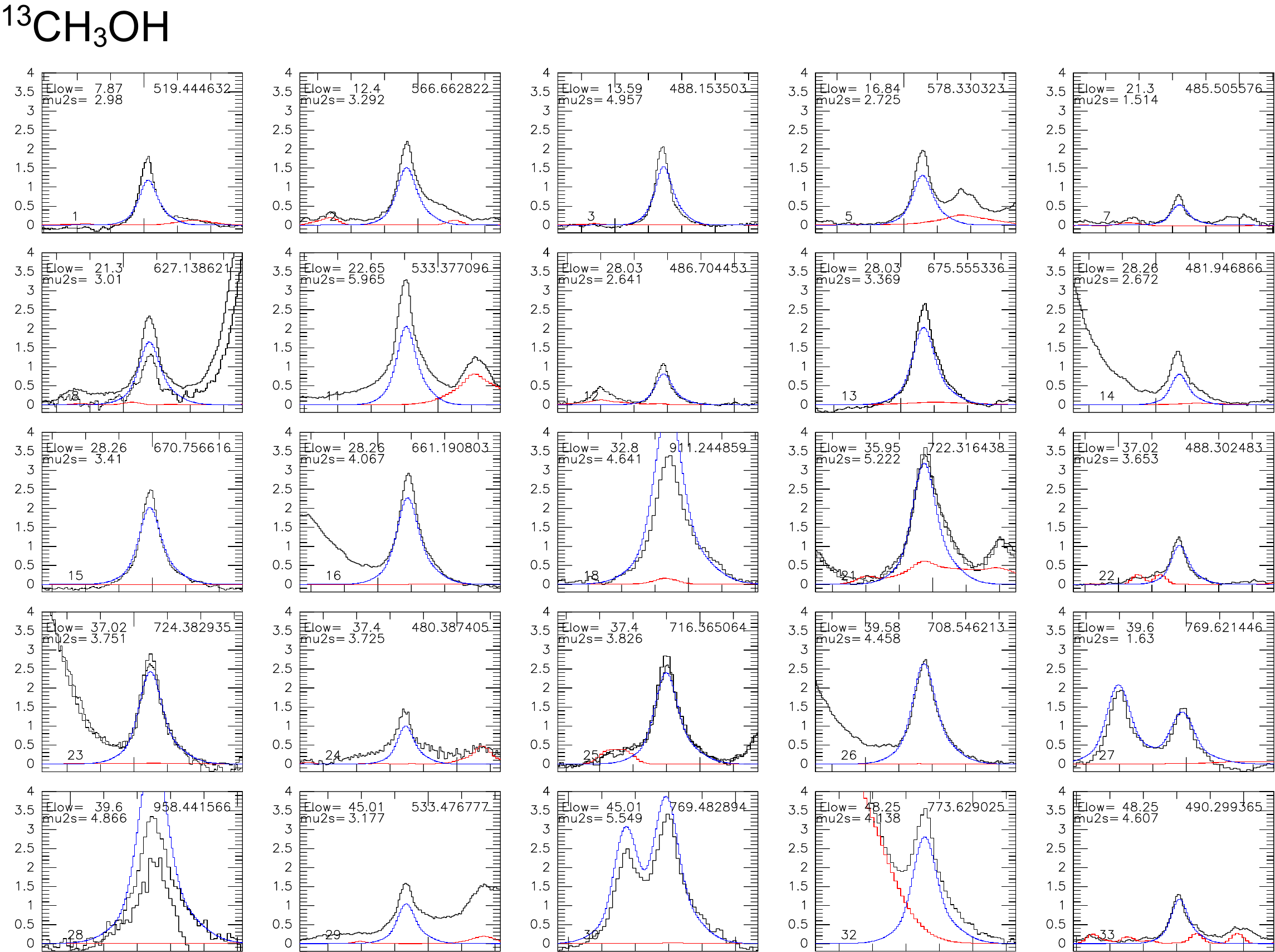}
\caption{Detected transitions of $^{13}$CH$_3$OH.  In each panel, the blue curve is the model to the emission of $^{13}$CH$_3$OH, and the red curve is the fullband model to all other species detected in Orion KL (Crockett et al. 2013b, in preparation).  The lower-state energy (in K), line strength (in D$^2$) and frequency (in GHz) is given in each panel.  The full Figure Set (10.1--10.21) can be found in the online form of the Journal. Each panel has a frequency width of 60 MHz.}
\end{figure}

\begin{figure}
\figurenum{11}
\includegraphics[width=6.0in]{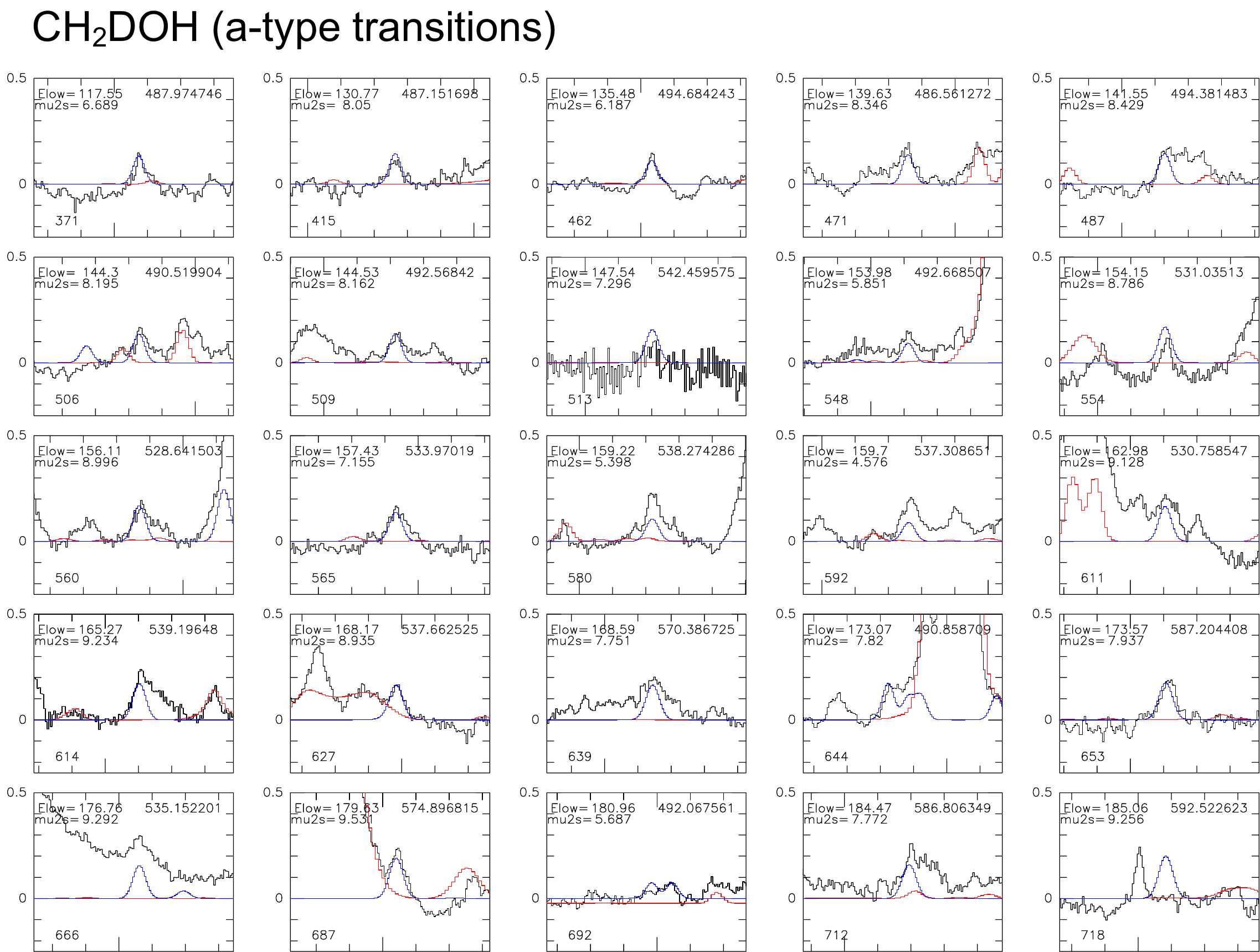}
\caption{Detected transitions of CH$_2$DOH.  In each panel, the blue curve is the model to the emission of CH$_2$DOH, and the red curve is the fullband model to all other species detected in Orion KL (Crockett et al. 2013b, in preparation).  The lower-state energy (in K), line strength (in D$^2$) and frequency (in GHz) is given in each panel.  The full Figure Set can be found in the online form of the Journal:  figures 11.1--11.5 show the $a$-type transitions, while figures 11.6--11.19 show the $b$- and $c$-type transitions.  Each panel has a frequency width of 60 MHz.}
\end{figure}

\begin{figure}
\figurenum{12}
\includegraphics[width=6.0in]{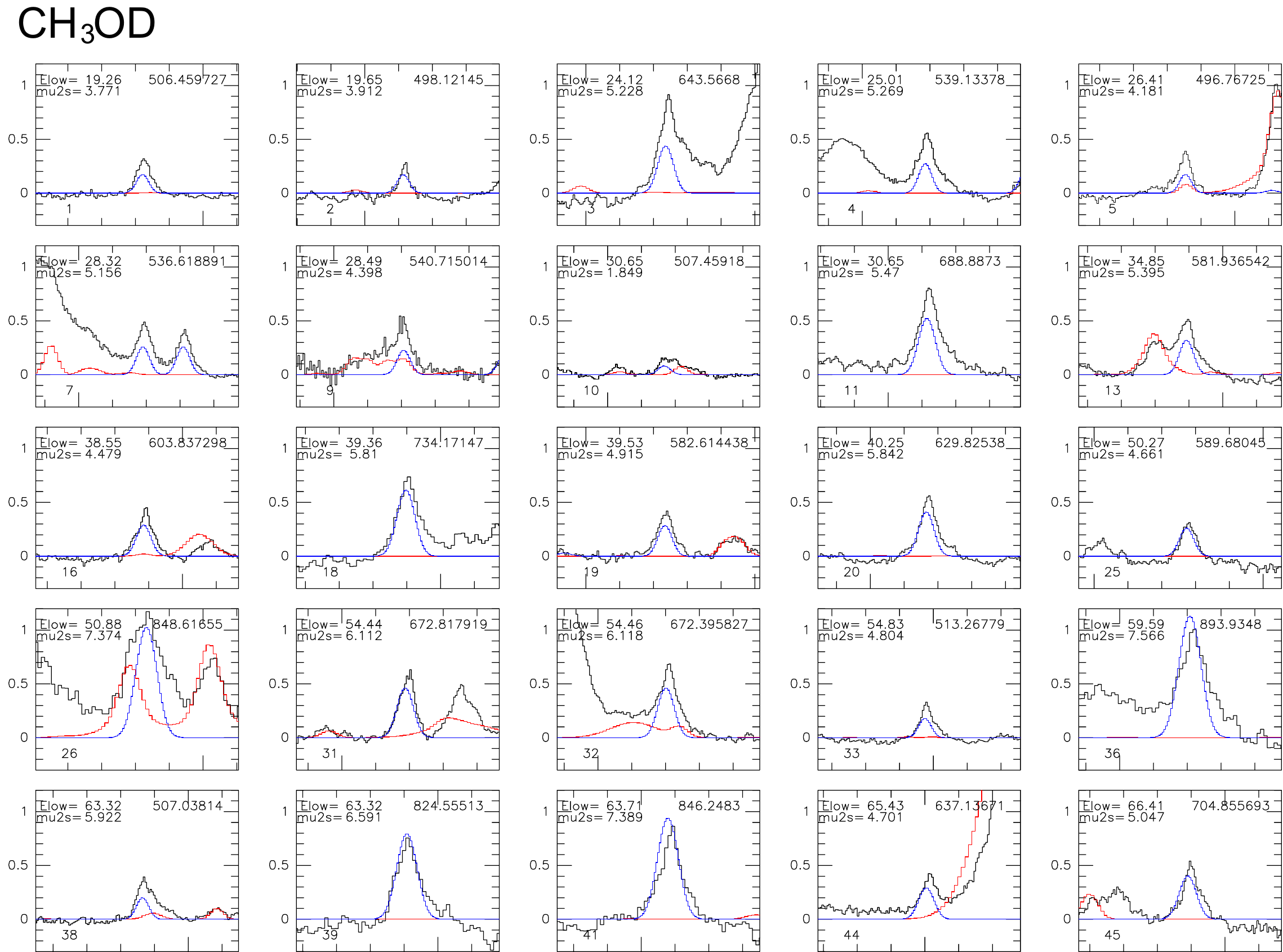}
\caption{Detected transitions of CH$_3$OD.  In each panel, the blue curve is the model to the emission of CH$_3$OD, and the red curve is the fullband model to all other species detected in Orion KL (Crockett et al. 2013b, in preparation).  The lower-state energy (in K), line strength (in D$^2$) and frequency (in GHz) is given in each panel.  The full Figure Set (12.1--12.12) can be found in the online form of the Journal.  Each panel has a frequency width of 60 MHz.}
\end{figure}

\end{document}